\documentclass{aa}  
\usepackage{graphicx}
\usepackage{txfonts}
\usepackage{natbib}
\bibpunct{(}{)}{;}{a}{}{,} 

\def\aaps{A\&AS}
\def\aap{A\&A}
\def\apj{ApJ}
\def\apjs{ApJS}

\newcommand{\hi}{{H\sc{i}~}}

\def\kms{km\,s$^{-1}$}
\def\kmss{km\,s$^{-1}~$}

\def\deg{\hbox{$^\circ$}}
\def\arcmin{\hbox{$^\prime$}}

\def\fdg{\hbox{$.\!\!^\circ$}}
\def\farcm{\hbox{$.\mkern-4mu^\prime$}}

\begin{document}

   \title{Anisotropies in the \hi gas distribution toward 3C~196}

   \subtitle{}

\author{P.\ M.\ W.\ Kalberla 
        \and J. Kerp         
}

\institute{Argelander-Institut f\"ur Astronomie, Universit\"at Bonn,
           Auf   dem H\"ugel 71, 53121 Bonn, Germany,  
           \\ 
  \email{pkalberla@astro.uni-bonn.de} }

   \authorrunning{P.\,M.\,W. Kalberla et al. } 

   \titlerunning{\hi anisotropies}

   \offprints{P.\,M.\,W. Kalberla}

   \date{Received 14 June 2016 / Accepted 18 August 2016 }

  \abstract 
  {The local Galactic \hi gas was found to contain cold neutral medium
    (CNM) filaments that are aligned with polarized dust emission. These
    filaments appear to be dominated by the magnetic field and in this
    case turbulence is expected to show distinct anisotropies. }
  {We use the Galactic Effelsberg--Bonn \hi Survey (EBHIS) to derive
   2D turbulence spectra for the \hi distribution in direction to 3C~196
   and two more comparison fields. }
 {Prior to Fourier transform we apply a rotational symmetric 50\% Tukey
   window to apodize the data. We derive average as well as
   position angle dependent power spectra. Anisotropies in the power
   distribution are defined as the ratio of the spectral power in
   orthogonal directions. }
 {We find strong anisotropies. For a narrow range in position angle, in
   direction perpendicular to the filaments and the magnetic field, the
   spectral power is on average more than an order of magnitude larger
   than parallel. In the most extreme case the anisotropy reaches
   locally a factor of 130. Anisotropies increase on average with
   spatial frequency as predicted by Goldreich and Sridhar, at the same
   time the Kolmogorov spectral index remains almost unchanged. The
   strongest anisotropies are observable for a narrow range in velocity
   and decay with a power law index close to --8/3, almost identical to
   the average isotropic spectral index of $-2.9 < \gamma < -2.6$. }
 {\hi filaments, associated with linear polarization structures in LOFAR
   observations in direction to 3C~196, show turbulence spectra with
   marked anisotropies. Decaying anisotropies appear to indicate that we
   witness an ongoing shock passing the \hi and affecting the observed
   Faraday depth.}

  \maketitle
%

\section{Introduction}
\label{intro}

The basic theory of a turbulent interstellar medium (ISM) with the
concept of the hierarchy of eddies was outlined by
\citet{vonWeizsaecker1951}. The energy flow from large eddies to smaller
ones was described in terms of wave numbers $k$, the reciprocal length
of the associated Fourier component, giving rise to the well known
Kolmogorov relation for the power distribution $ P(k) \propto k^{-5/3}$.
In a more general way for the decay of turbulent energy within a
compressible medium a constant power law index $ P(k) \propto
k^{-\gamma}$ is expected. When the turbulent flow reaches the
dissipative range where turbulent motion is converted to heat, the power
distribution should steepen strongly, $ P(k) \propto k^{-7}$.

The measurement of $ P(k)$ is, in principle, straightforward for radio
synthesis telescopes by using the squares of the observed visibility
amplitudes at various baselines. In practice, however, a number of
important details such as calibration, noise performance, UV plane
coverage (sampling bias) and projection effects need to be taken into
account \citep{Green1993}. Most important is to apply a uniform taper in
the UV plane which is usually avoided when generating sensitive
interferometer maps \citep{Briggs1995}. Inappropriate tapering may
degrade power spectra seriously.

Single-dish radio telescopes can be used to generate 1D power spectra
from drift scans or average power spectra from 2D maps. Tapering issues,
discussed later in more detail (Sect. \ref{Apodisation}), are also in
this case important. Essentially 2D single-dish data have to be
processed analogous to interferometer data for an accurate power
spectrum analysis; instrumental effects need to be eliminated carefully.

Attempts to quantify ISM power law indices for a turbulent \hi
distribution were made by several authors. In the first instance 2D
interferometer channel maps, slices of the observed brightness
temperature distribution at a constant velocity, were
used. \citet{Crovisier1983} found power law spectra that were
independent of position angle with typical values of $\gamma \sim -3$
for Westerbork synthesis telescope observations. From additional 1D
Arecibo and Nan\c{c}ay drift scans they derived $\gamma \sim
-2$. Erroneously it was concluded that interferometer and single-dish
telescopes sample \hi distributions with different properties. This
issue was solved later, as it was realized that power laws derived from
1D and 2D distributions result in power law indices differing by one
\citep{Green1993}.

\citet{Kalberla1983} used also Westerkork observations, supplemented by
short spacing data from the Effelsberg 100-m telescope, and derived for
local \hi gas toward 3C147 $\gamma = -2.5 \pm 0.3$.  Using the DRAO
Galactic Plane Survey for \hi emission, \citet{Green1993} derived power
law indices between $ -3 < \gamma < -2.2 $. More distant \hi gas, which
corresponds to large physical scales, tends to have more negative
indices than the nearer \hi. 

After these early investigations it became obvious that turbulence in
the \hi gas cannot be described by a unique power law. Rather turbulence
in the \hi is governed by processes that inject the kinetic energy to
the ISM with different decay processes.  In fact, the power spectrum
of a velocity channel is a complex mixture of velocity and density
fluctuations. \citet{Lazarian2000} studied these issues in detail.  They
argued that intensity fluctuations within individual channel maps are
generated predominantly by turbulent velocity fields. Increasing the
thickness of the observed velocity slices causes density fluctuations to
dominate the emissivity. For thin slices in velocity preferentially a
power law index close to $\gamma = -8/3$, the 2D Kolmogorov index,
is expected. The analysis of column density maps (thick slices in
velocity) should result in power law indices that are steeper by one
\citep[][Eq. 28]{Lazarian2000}.  \citet{Esquivel2005} argued that
velocity centroids are most useful as tracers of turbulence and favor
centroid maps for anisotropy studies.

Using data from the Southern Galactic Plane Survey in the fourth
Galactic quadrant, \citet{Dickey2001} derived turbulence spectra with
power law indices between $ -4 < \gamma < -3 $, confirming essentially
the predicted steepening for column density maps.
\citet{Miville-Deschenes2003} studied the Ursa Major Galactic cirrus and
found that the spectral index is similar for the 3D density and velocity
fields with a value of $ -3.6 \pm 0.2$. Also
\citet{Miville-Deschenes2007} found that the power spectra of the
integrated emission and velocity centroids are both well fit by a
slope of $-3.4 \pm 0.2$. Using Arecibo data \citet{Dedes2012} report for
thin slices a spectral index of -2.67 for the local gas and -2.5 for the
extra-planar component.  \citet{Martin2015}, using the Green Bank
Telescope (GBT), studied the \hi distribution in toward the northern
ecliptic pole. They derived power spectra of column density maps for
low-, intermediate-, and high-velocity gas components (LVC, IVC, and
HVC) with indices of $-2.86 \pm 0.04, -2.69 \pm 0.04$, and $-2.59 \pm
0.07$, respectively.  \citet{Blagrave2016} observed several fields at
intermediate Galactic latitudes with the DRAO and the GBT and find
exponents ranging from -2.5 to -3.0. Power spectra of maps of the
centroid velocity for these components give similar results.

For the Small Magellanic Cloud (SMC) \citet{Stanimirovic1999} derived
power law indices in the range $\gamma = -3.04 \pm 0.02 $. This
implies that the dynamics of the \hi distribution are similar for both
the Galaxy and the SMC.

All these investigation are consistent with the assumption that the ISM
is dominated by turbulence. Since a few years increasingly attention
was drawn to the fact that typical astrophysical fluids are accompanied
by magnetic fields. These fields may be strong enough to affect the
dynamics of the turbulent cascade from large to small eddies. The motion
along the magnetic field is free and unaffected, but perpendicular to
the field motions are hindered. During the energy cascade this leads
to increasingly asymmetric eddies. Thus magneto hydrodynamical (MHD)
turbulence in presence of a magnetic field is expected to be strongly
anisotropic \citep{Goldreich1995}, for reviews see
\citet{Cho2003,Brandenburg2013}.

So far, imprints to the spectral power distribution from anisotropies in
the \hi distribution were not reported. However recently new powerful
single-dish \hi surveys became available. For the northern sky there is
the Galactic Effelsberg--Bonn \hi Survey \citep[EBHIS,][]{Winkel2016a},
in the south the Parkes Galactic All Sky Survey
\citep[GASS,][]{Naomi2009,Kalberla2010}. \citet{Kalberla2016}, merging
EBHIS with the 3rd release of the GASS \citep[GASS III,][]{Kalberla2015}, have
shown that the cold neutral medium (CNM) in the local vicinity is mostly
organized in filaments. A comparison with structures measured by {\it
  Planck} at 353 GHz \citep{Planck2016} disclosed that the CNM filaments
are well aligned with the thermal dust emission. Moreover, the \hi
filaments are well aligned with the magnetic field direction measured by
{\it Planck} and anisotropies in the CNM distribution are common. We
therefore find that a search for anisotropic MHD turbulence is overdue
and decided to dedicate this analysis to anisotropies in \hi turbulence
spectra.

\begin{figure}[tbp]
   \centering
   \includegraphics[width=6.5cm,angle=-90]{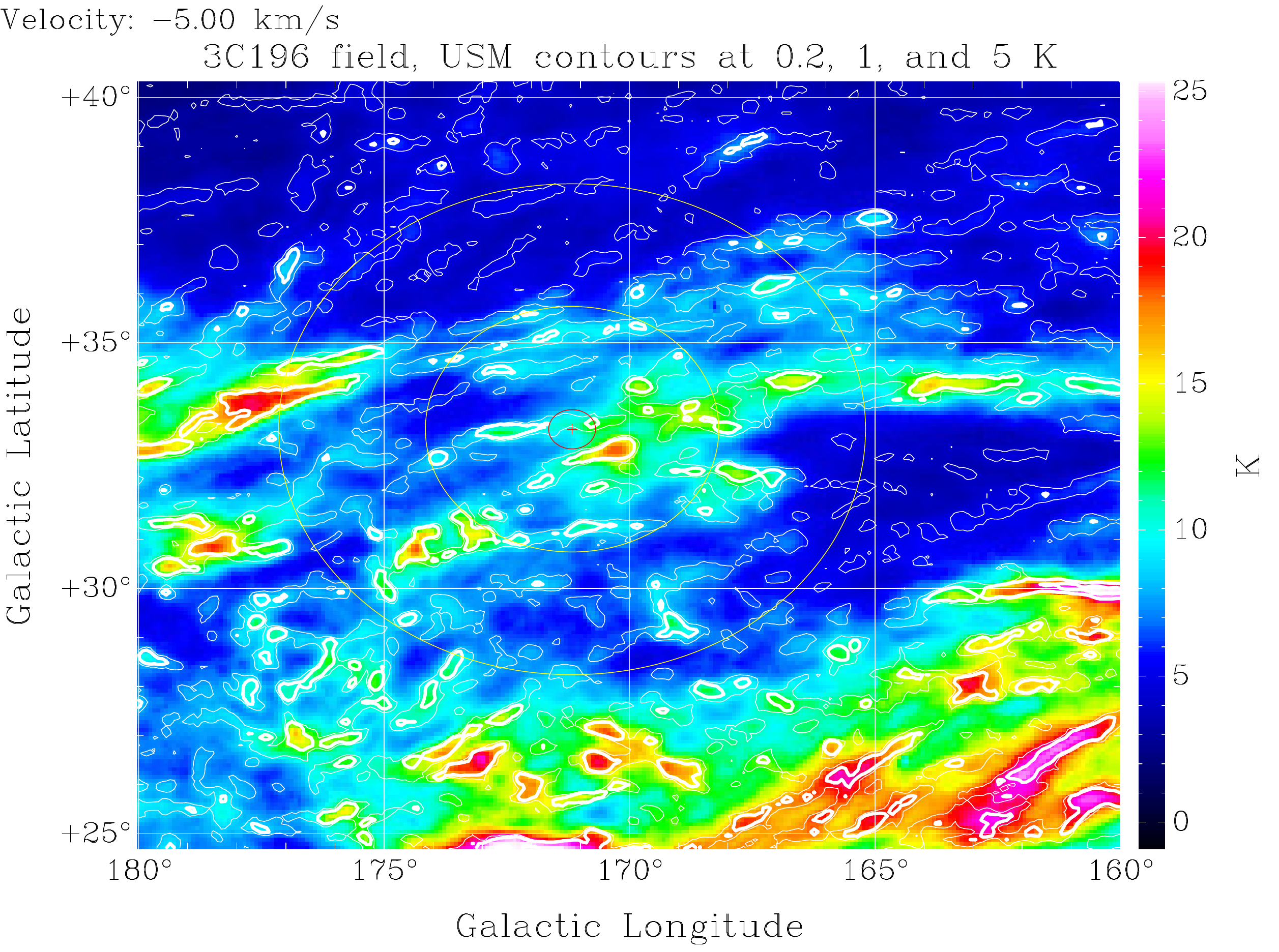}
   \caption{Brightness temperature distribution in direction to 3C~196
     in Galactic coordinates at a velocity of -5 \kmss on large
     scales. 3C~196 is marked in red, the yellow circles have radii of
     2.5\degr and 5\degr. Filamentary features from USM maps are
     overlaid with contours of 0.2, 1, and 5 K.   }
   \label{Fig_HI_gal}
\end{figure}

\begin{figure}[tbp]
   \centering
   \includegraphics[width=6.5cm,angle=-90]{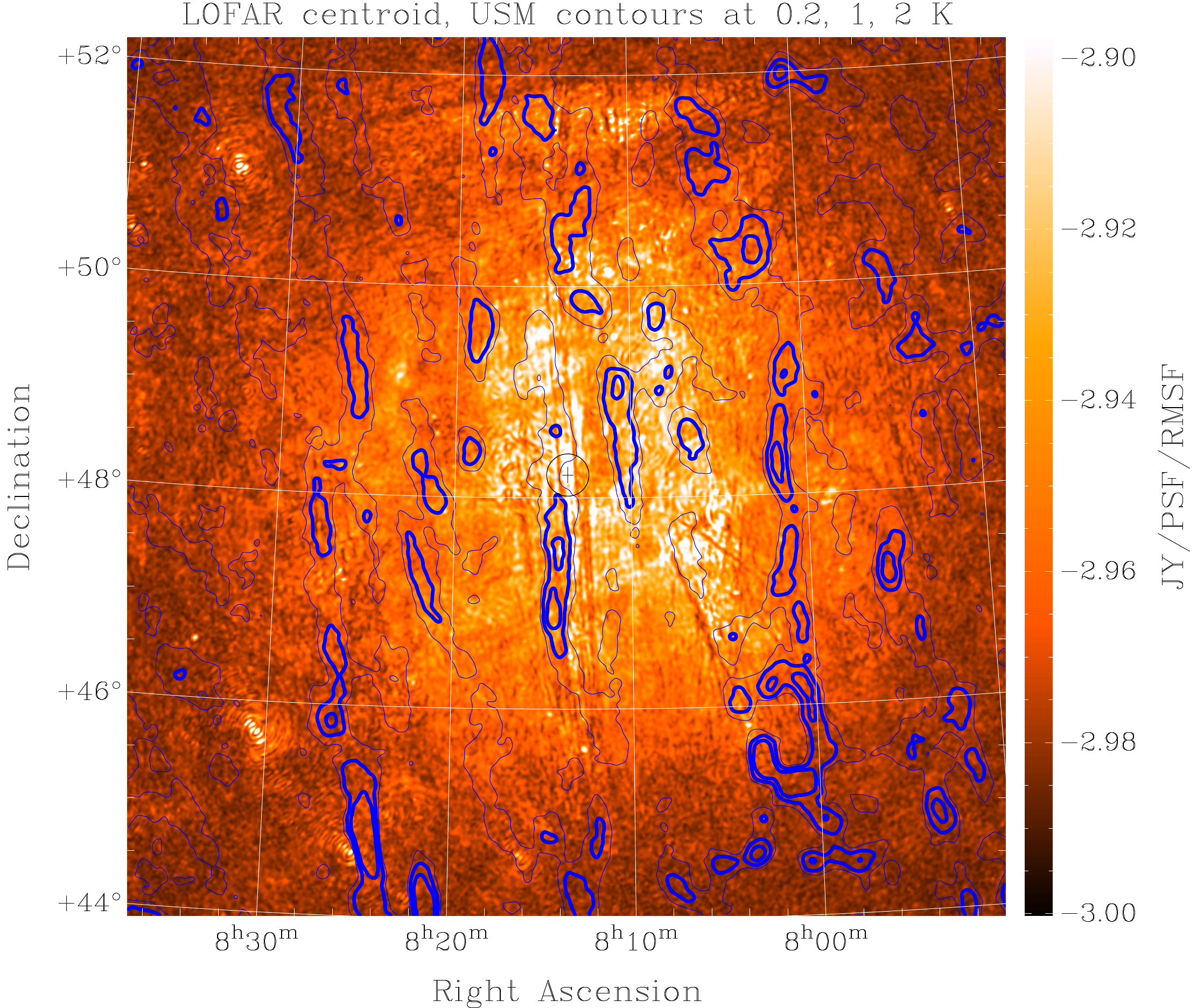}
   \caption{Centroid map (Faraday depth in rad m$^{-2}$) in ecliptic
     coordinates derived from the LOFAR polarized emission in direction
     to 3C~196. The filamentary features from an
     \hi USM map at $v_{\rm LSR} = -3.76$ \kmss are overlaid with
     contours of 0.2, 1, and 2 K. The position of 3C~196 in the field
     center is indicated. }
   \label{Fig_cubep_USM}
\end{figure}

\section{Source selection}
\label{source}

Linear polarization structures in LOFAR observations of the interstellar
medium in the 3C~196 field have been observed by \citet{Jelic2015}. This
is one of the three primary fields of the LOFAR-Epoch of Reionization
key science project. The high band antennas (HBA) have been used to
image this region and the distribution of polarized structures in
Faraday depth was derived by rotation measure (RM) synthesis.

These observations disclosed an interesting morphological
feature, a strikingly straight filament at a Faraday depth of +0.5 rad
m$^{-2}$, approximately in north/south direction in the center across
3C~196, and parallel to the Galactic plane. The width of the central
filament is roughly 8 to 10 arcmin.  Moreover, there are also linear
depolarization canals conspicuous in images at the peaks of the Faraday
spectra. \citet{Jelic2015} conclude that the filamentary structure is
probably caused by an ionized filament in the ISM, located somewhere in
the foreground.

We used the EBHIS to study the \hi distribution
in direction to 3C~196. The LOFAR field field is located in a region at
intermediate Galactic latitudes that is dominated by numerous \hi
filaments, running approximately parallel to the Galactic plane. The
large scale structure of the filaments is best visualized at a velocity
of -5 \kms.  Figure \ref{Fig_HI_gal} displays color coded an overview of
the observed brightness temperature $T_{\rm B}$. Filaments are best
visible after unsharp masking (USM) the observations. We generated USM
maps by subtracting from the observed $T_{\rm B}$ distribution a
smoothed distribution with an effective resolution of 0\fdg5
\citep[see][for details]{Kalberla2016}. For the local gas the resulting
filamentary structures are found to have low Doppler temperatures,
characteristic for the CNM. This gas is shown in Fig. \ref{Fig_HI_gal}
with isophotes. These structures are typical for the filamentary CNM
described by \citet{Kalberla2016}.

For a more detailed comparison between \hi filaments and the polarized
emission we calculate the centroid map from the LOFAR data cube of
polarized emission at various Faraday depths. Such a centroid map should
be sensitive to shifts in Faraday depth, caused by foreground
filaments. In Fig. \ref{Fig_cubep_USM} we display this centroid map,
overlaid with CNM filaments (red contours) from the EBHIS \hi USM
database at a velocity of $v_{\rm LSR} = -3.76$ \kms. We find most of
the \hi filaments aligned with dips (dark blue) in the LOFAR centroid
map but there is no unambiguous one-to-one correlation, even when
checking the CNM at other velocities. However a comparison between \hi
gas and LOFAR data shows in general a good alignment between CNM
filaments and polarized structures in Faraday depth. Moreover,
\citet{Zaroubi2015} find a very clear correlation between the
filamentary structures, detected with LOFAR, and the magnetic field
orientation, probed by the {\it Planck} satellite. This is consistent
with \citet{Kalberla2016} who report that there is in general a good
agreement between CNM filaments and the magnetic field direction, probed
by {\it Planck}.

After comparing LOFAR, EBHIS and {\it Planck} data at 353 GHz we suspect
that LOFAR filaments in polarized emission and \hi are related to each
other. Both probe the magnetic field as observed by {\it Planck}.  We
consider therefore the 3C~196 field as a prime target to study the
relations between magnetic field orientation and induced anisotropic
turbulence in the \hi gas.

\section{Preparing the data for analysis }
\label{data_analysis}

\subsection{Database}
\label{Data_base}

We use data from the first EBHIS data release \citep{Winkel2016a}. This
survey covers all declination $\delta > -5\degr $. The rms uncertainties
of the brightness temperatures per channel are 90\,mK at a full width at
half maximum (FWHM) width of $\delta v = 1.44$ \kms. From the original
EBHIS database, corrected for instrumental baselines, radio
interference and stray radiation, we generated FITS data cubes, covering
a region of 10\degr x 10\degr~ in equatorial coordinates with a
rotational symmetric Gaussian beam of 10\farcm8 FWHM beam-size
\citep{Winkel2016b}.

\subsection{Spectral analysis}
\label{Apodisation}

To quantify anisotropies in the spectral power distribution it is
necessary to derive unbiased angular dependencies.  Calculating a map of
the \hi distribution demands restricting the range of the source
distribution on the sphere by applying a window function. After Fourier
transform one obtains a convolution of the transformed true distribution
with the transform of the window function \citep{Bracewell2000}. For a
rectangular window this causes a cross-like structure in the UV plane,
with a particular nasty strong response at low spatial frequencies. To
mitigate this effect the image data are usually apodized at the field
boundaries with a $\sim 5$ pixel wide cosine function, afterwards the
median is used to calculate the power spectrum, avoiding strong biases
from the cross \citep[e.g.,][]{Martin2015}.

This recipe mitigates effects from the window response in case of
average power spectra. When deriving an angular power distribution we
found that the artificial cross structure is still dominant. To overcome
this bias, a strictly spherical apodization with a strong taper is
needed. It is further desirable to use a large untapered region but with
low sidelobes of the window function in the Fourier domain. After some
tests we have chosen a rotational symmetric 50\% cosine taper (Tukey)
window \citep{Harris1978}, leaving the central 50\% of the image
untapered and applying then a cosine taper that reaches zero at the
field boundary. In the UV plane this taper function leads to a slight
smoothing, similar to the smoothing caused by the Fourier transform of
the primary beam function in case of interferometer observations.  The
first sidelobe of the Fourier transformed Tukey taper is at a level of
-20 dB \citep[see Fig. 31,][]{Harris1978}.

\begin{figure*}[tbp]
   \centering
   \includegraphics[width=6.0cm,angle=-90]{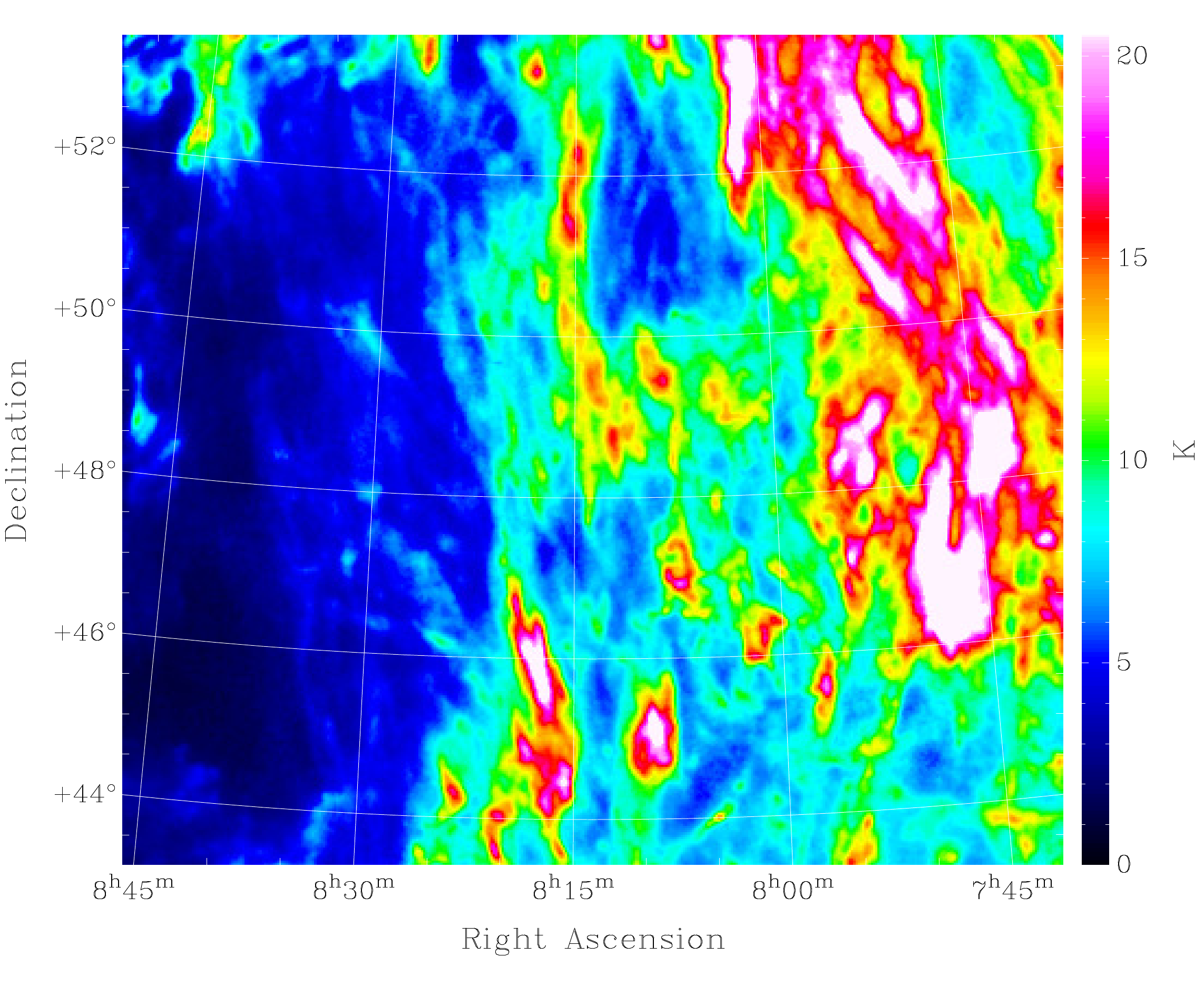}
   \includegraphics[width=6.0cm,angle=-90]{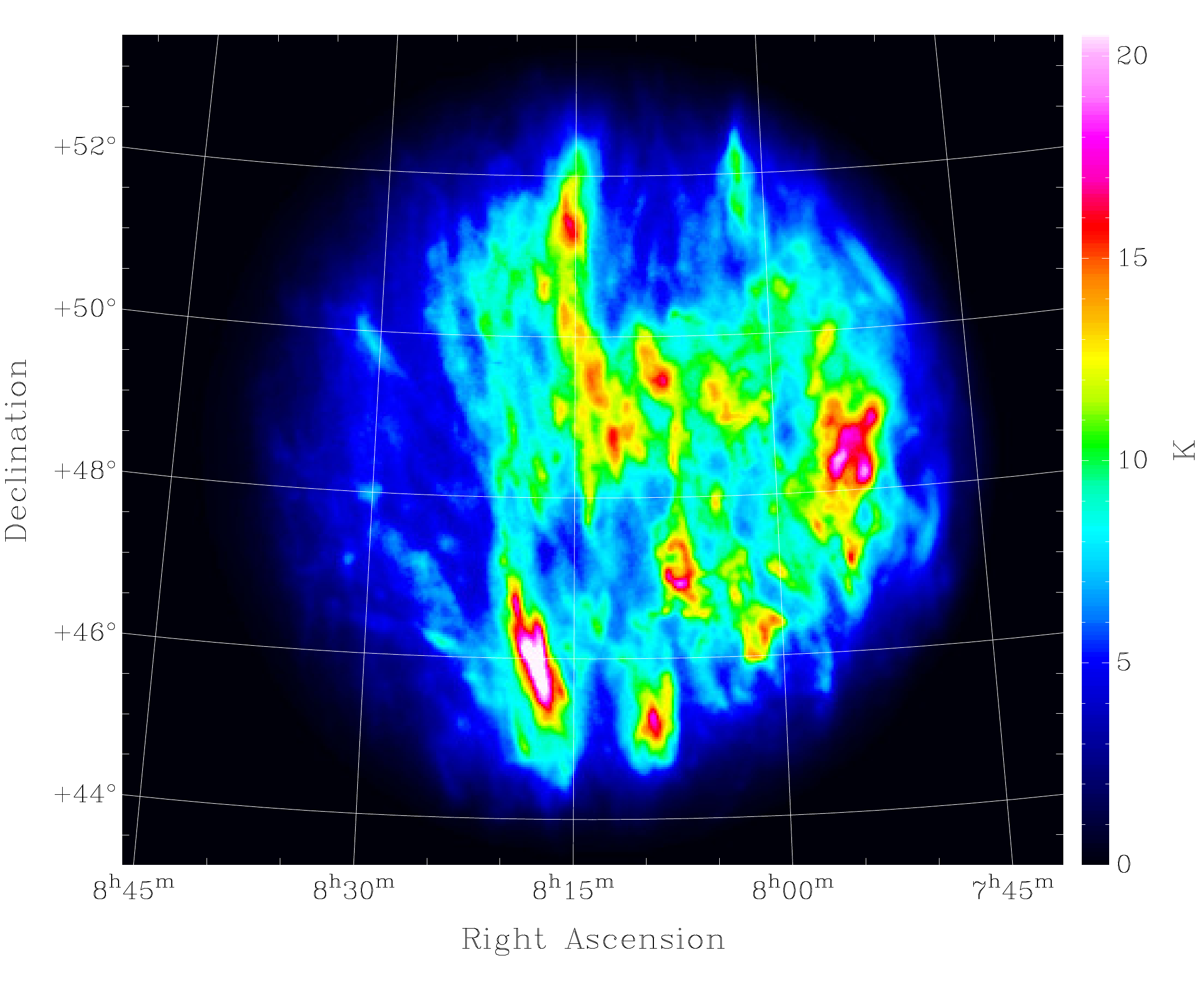}
   \caption{{\it left:} \hi $T_{\rm B}$ distribution at $v_{\rm LSR} = -3.76$
     \kmss as observed, {\it right:} \hi $T_{\rm B}$ distribution after
     apodization with a 50\% Tukey taper. }
   \label{Fig_image_37}
\end{figure*}

\begin{figure*}[htb]
   \centering
   \includegraphics[width=6.3cm,angle=-90]{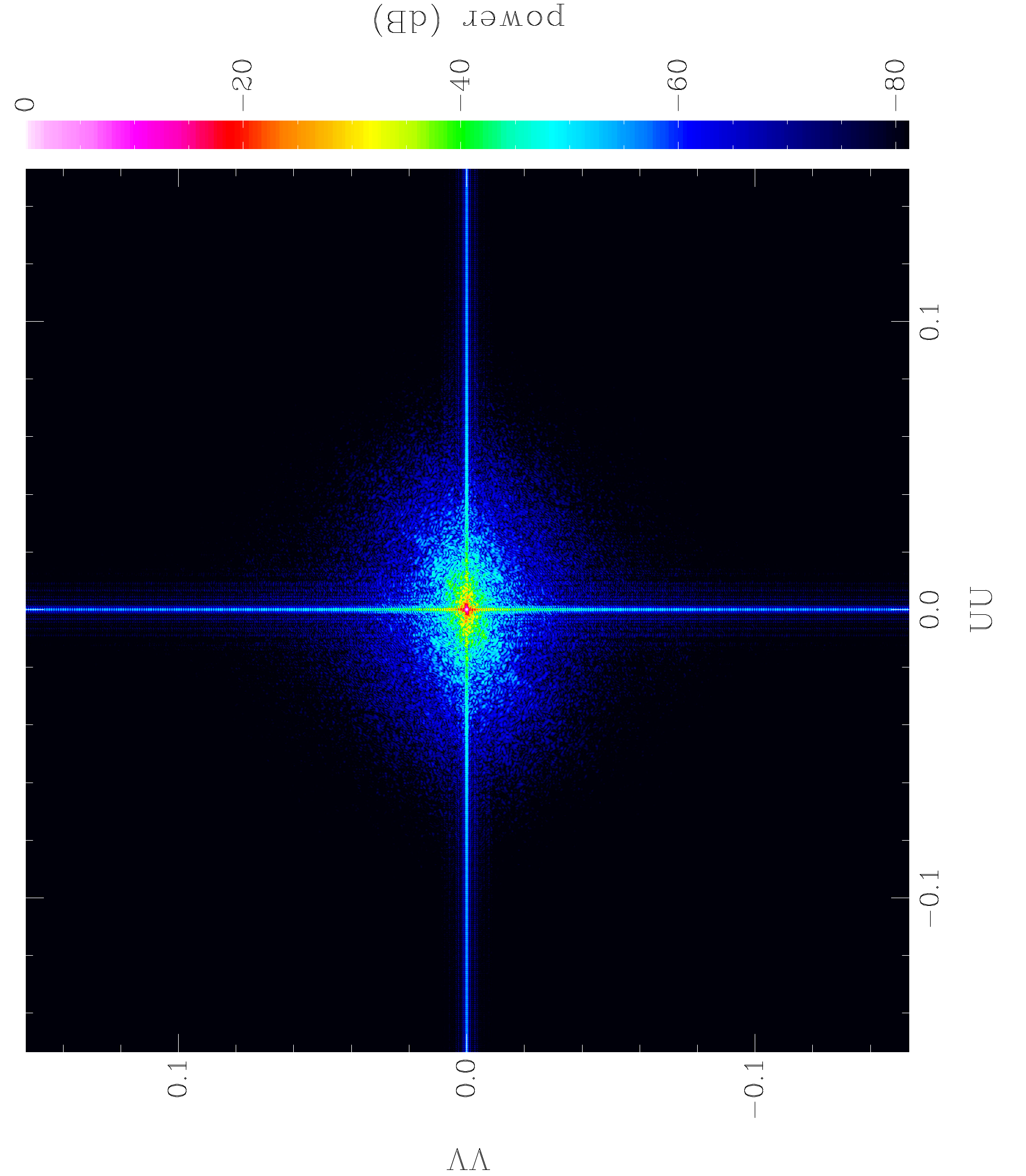}
   \includegraphics[width=6.3cm,angle=-90]{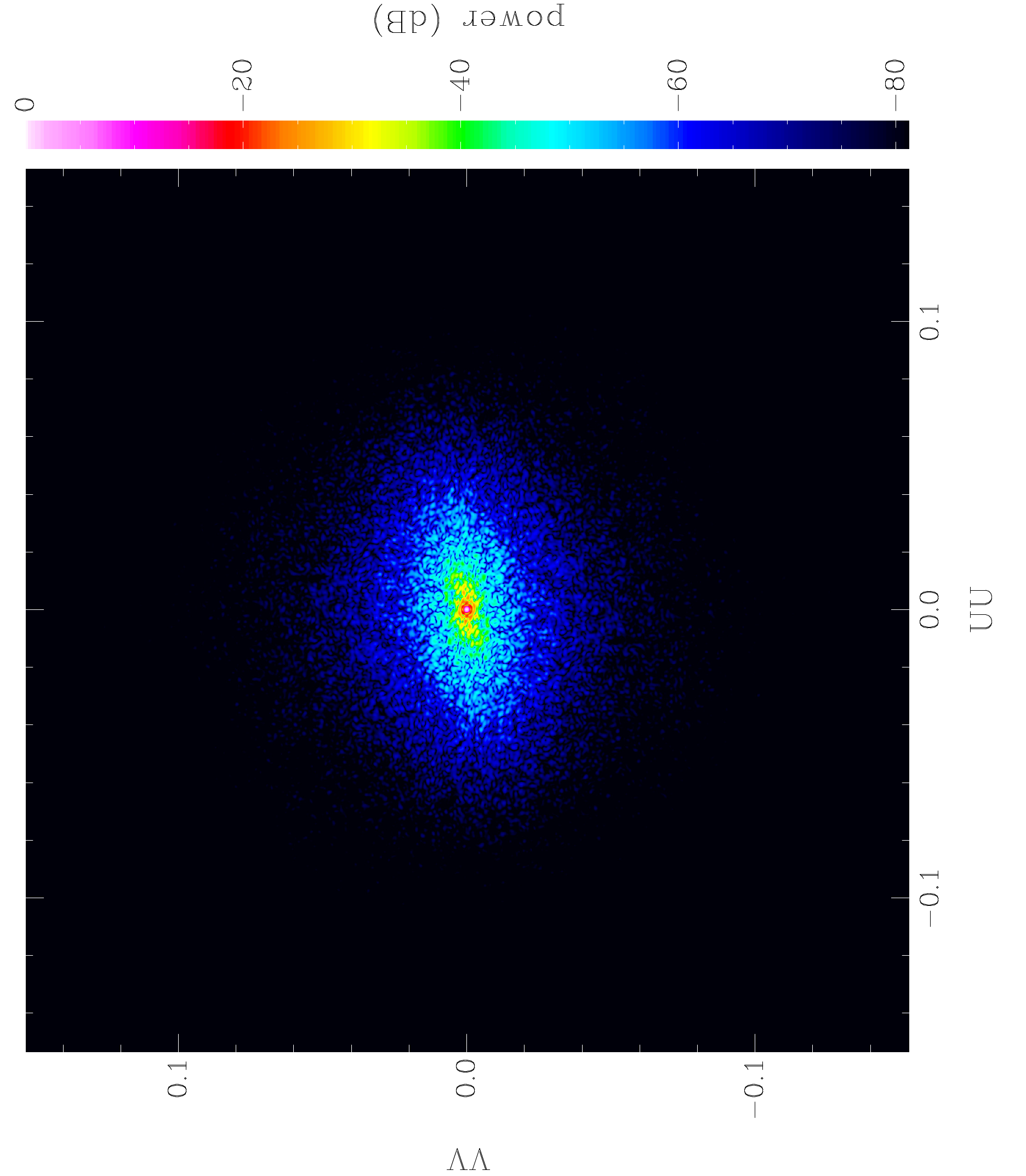}
   \caption{{\it left:} normalized power distribution at $v_{\rm LSR} = -3.76$ 
     \kmss as observed, {\it right:} \hi power distribution 
     of the apodized \hi map. The scales are logarithmic in dB. }
   \label{Fig_power_37}
\end{figure*}

In the following we demonstrate the individual steps is our data
processing. In Fig. \ref{Fig_HI_gal} we indicate with yellow circles the
inner untapered 50\% and the outer range (100\%) where the data are
tapered to zero.  In Fig. \ref{Fig_image_37} we display on the left side
the original image of the \hi distribution at $v_{\rm LSR} = -3.76$
\kms. On the right the same after apodization with a 50\% Tukey taper.
In analogy to a synthesis radio telescope, this image resembles a
primary beam attenuated image.  Figure \ref{Fig_power_37} displays the
power distribution, the squared moduli of the image after Fourier
transform. Here we used for display a logarithmic representation, at the
left without and at the right with apodization.

\begin{figure*}[htb]
   \centering
   \includegraphics[width=6.3cm,angle=-90]{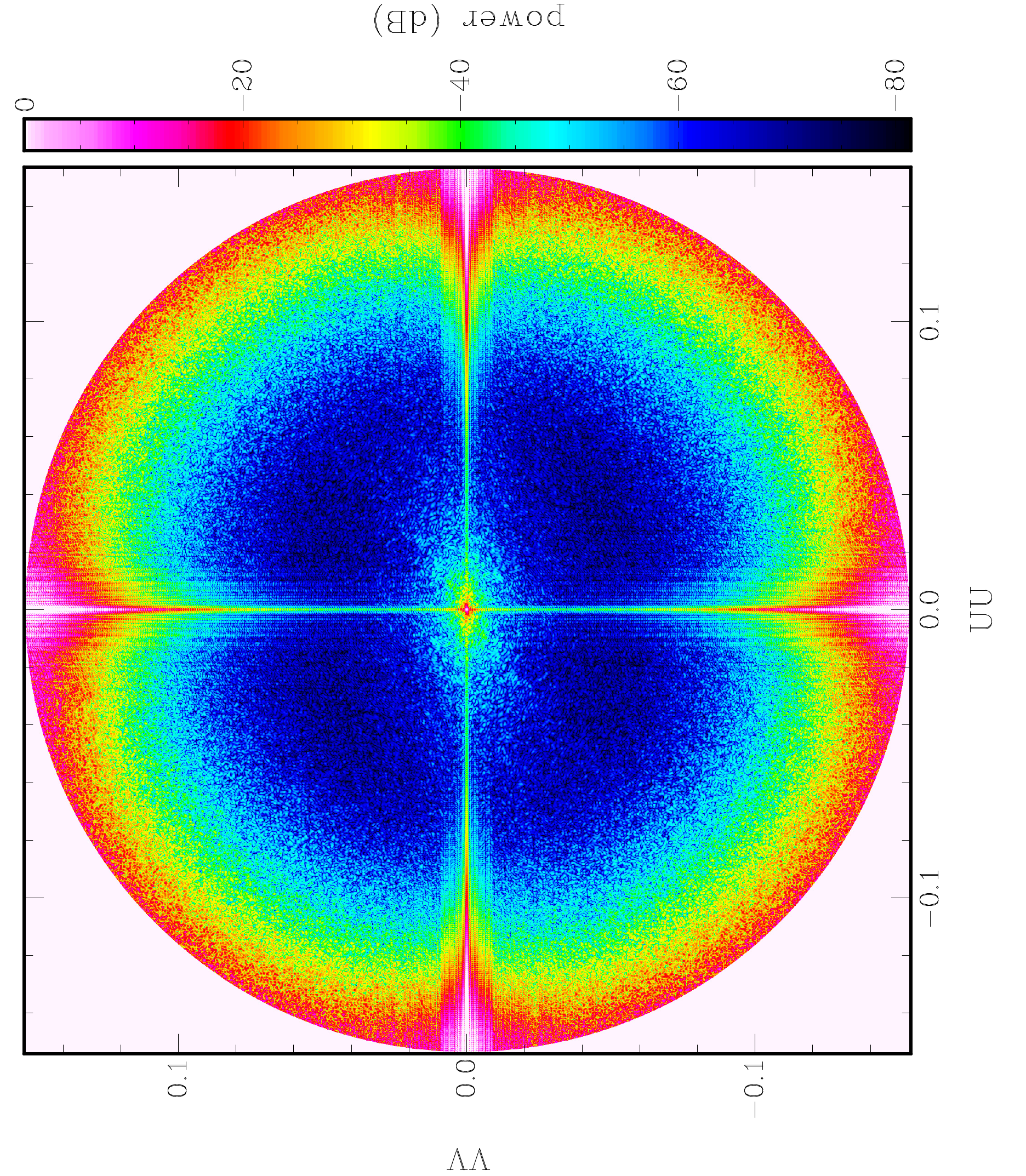}
   \includegraphics[width=6.3cm,angle=-90]{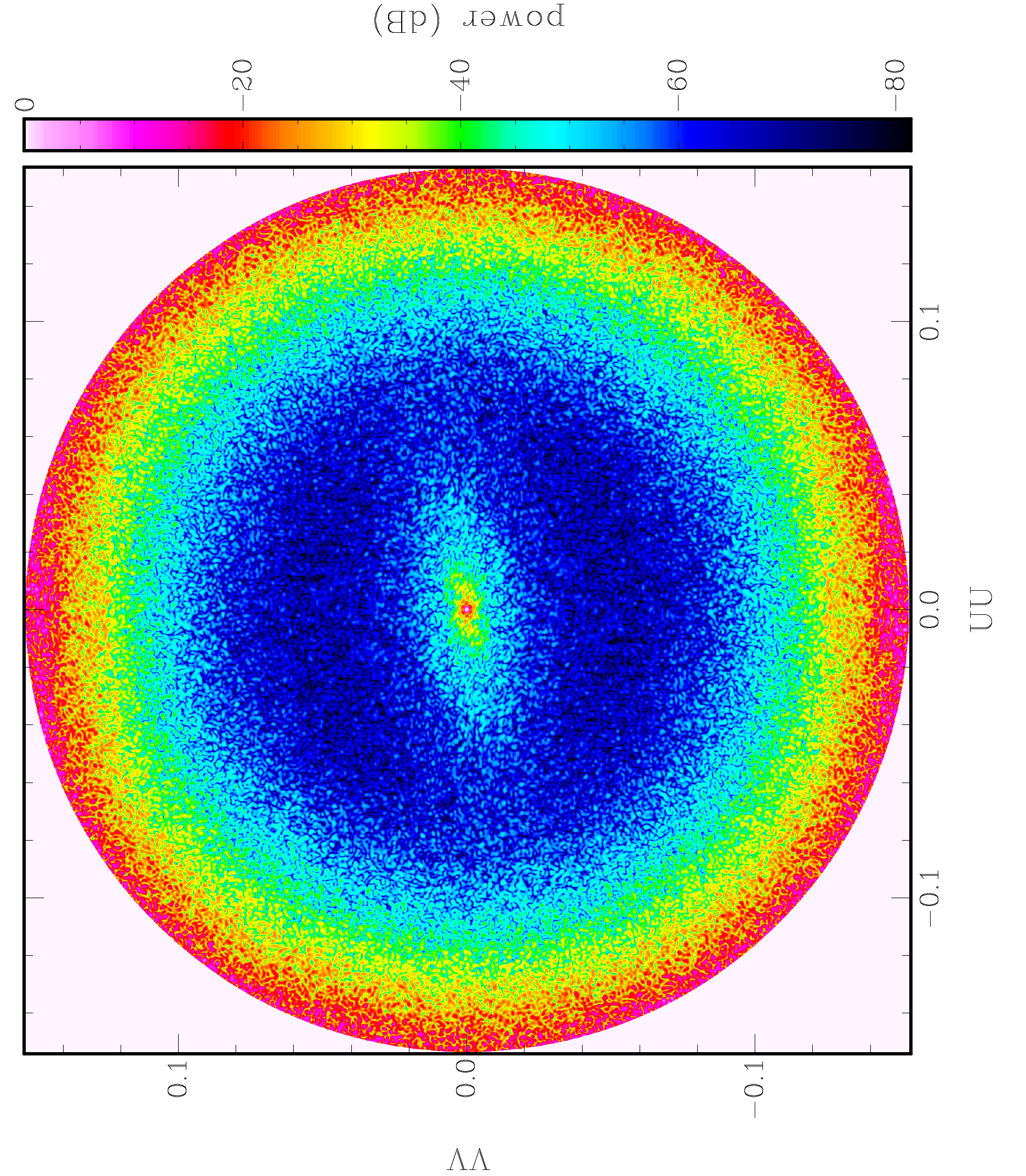}
   \caption{{\it left:} normalized power distribution at $v_{\rm LSR} =
     -3.76$ \kms, beam corrected, {\it right:} \hi power distribution of
     the beam corrected apodized \hi map. The scales are logarithmic in
     dB.  }
   \label{Fig_power_37_correct}
\end{figure*}

The EBHIS maps are gridded with a Gaussian kernel. Data processing as
well as smoothing due to the telescope beam result in an effective
Gaussian beam of 10\farcm8. A deconvolution is thus necessary. After
Fourier transform we divide the derived 2D power distribution by the
square of the transformed beam response.  The result is shown in
Fig. \ref{Fig_power_37_correct}. On the right is the 2D power spectrum
that we use for further analysis. It is obvious from this Fig. that
high spatial frequencies are severely affected by instrumental noise
that was amplified by the beam response function. This contribution can
safely be considered as isotropic, without position angle dependent
biases.

\section{Deriving the spectral index}
\label{Deriving}

After generation of \hi maps, subsequent apodization and Fourier
transform with correction for beam convolution as explained in the
previous Sect., we proceed with the data analysis.

\begin{figure}[tbp]
   \centering
   \includegraphics[width=6.5cm,angle=-90]{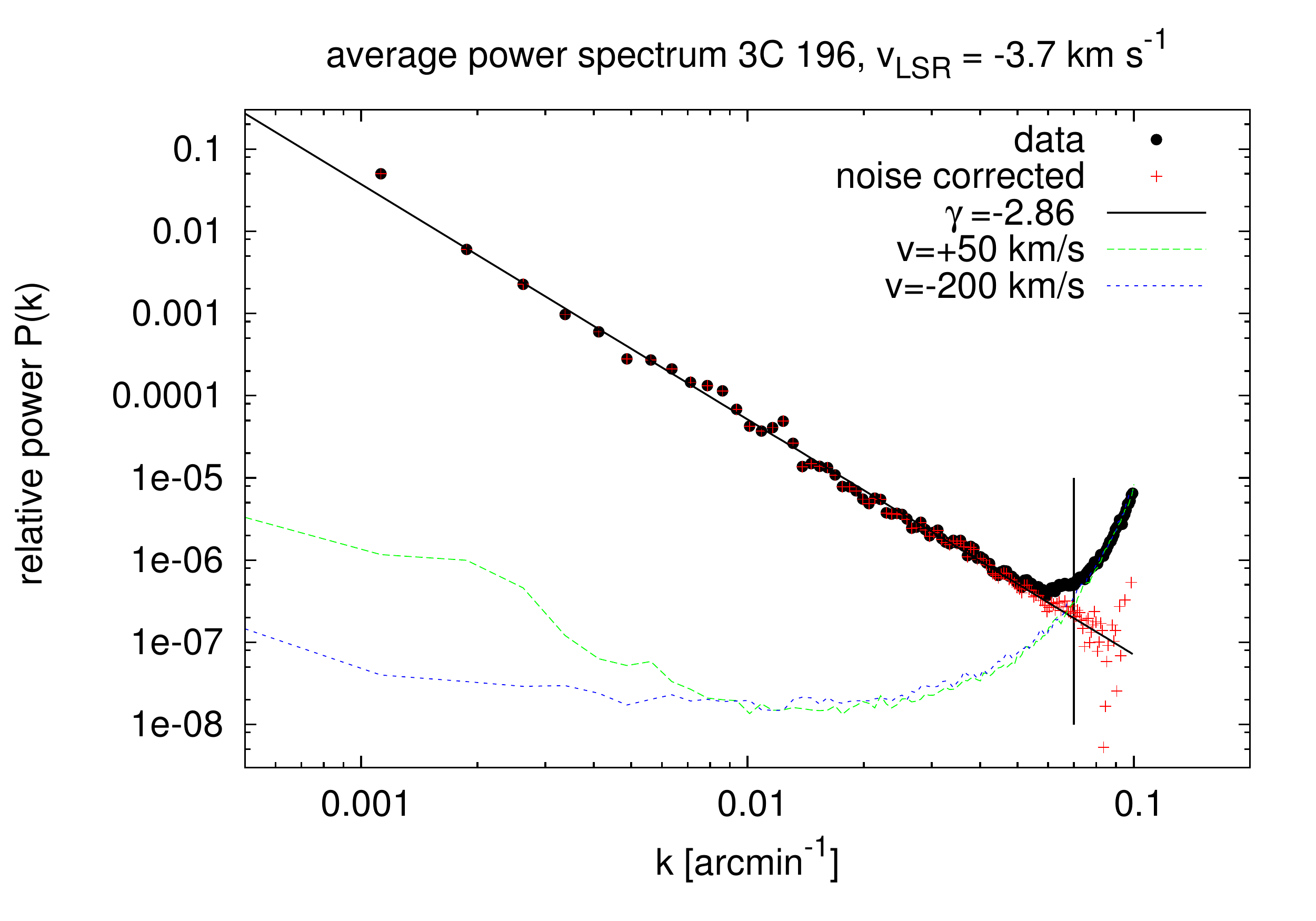}
   \caption{EBHIS averaged angular power spectrum (black dots) at
     $v_{\rm LSR} = -3.76\,{\rm km\,s^{-1}}$. Systematic and statistical
     uncertainties are estimated quantitatively by the power spectra
     calculated for emission free portion of the EBHIS data at $v_{\rm
       LSR} = +50\,{\rm km\,s^{-1}}$ (green) and $v_{\rm LSR} =
     -200\,{\rm km\,s^{-1}}$ (blue). Data after subtraction of a noise
     template $N(k)$ are plotted in red. The distribution for $ k <
     0.07$ arcmin$^{-1}$ (vertical line) is approximated by a power-law
     with $\gamma = -2.86\,\pm\,0.03$.  }
   \label{Fig_spec_aver}
\end{figure}

\subsection{Average power spectrum and noise}
\label{Averagepower}

To determine the average (isotropic) power spectrum we integrate the 2D
power distribution in annuli of constant spatial frequencies $ k =
(k^2_x + k^2_y)^{1/2} $. As an example we use here the data from
Fig. \ref{Fig_power_37_correct} (right panel) at $v_{\rm LSR} = -3.76$
\kms. The result is plotted in Fig. \ref{Fig_spec_aver} marked by black
dots. We obtain an average power distribution $P(k)$ that can be
described by the relation
\begin{equation}
P(k) = c \cdot k^\gamma + N(k).
\label{eq:Pav}
\end{equation}
Here $c$ is an arbitrary scale factor, resulting from the fact that we
normalize in all cases the total power of the 2D power maps.
$\gamma$ is the spectral index and $N(k)$ the contribution due to
instrumental noise. It is obvious from Fig. \ref{Fig_power_37_correct}
(right panel) that the noise dominates at high spatial frequencies. This
is also clearly visible in Fig. \ref{Fig_spec_aver}.

The noise contribution $N(k)$ to the power spectrum is in principle
unknown, but can easily be estimated from channel maps without \hi
emission. In Fig. \ref{Fig_spec_aver} we plotted as examples the noise
power measured at $v_{\rm LSR} = +50 $ \kmss and $v_{\rm LSR} = -200$
\kmss after matching the power spectra at high spatial frequencies. A
slight matching of the noise power is necessary because the system noise
is not constant across the observed band, also the system noise
increases with line temperature.

The observations suffer from uncertainties caused by radio frequency
interference (RFI) and residual baseline uncertainties, including
remaining errors in the correction for stray radiation, see
\citet{Winkel2016a} for discussion. The noise spectra in
Fig. \ref{Fig_spec_aver} differ predominantly for $ k < 0.008 $
arcmin$^{-1}$, corresponding to scales larger than 2\degr\ and are
caused by such systematic uncertainties. The two examples in
Fig. \ref{Fig_spec_aver} represent approximately the range of expected
uncertainties for EBHIS observations. Using 40 emission free
channels we derive a noise template $N(k)$.  Subtracting $N(k)$ we
derive the unbiased normalized power spectrum $P_{\rm av}(k) =  k^\gamma $,
plotted in Fig. \ref{Fig_spec_aver} in red, and fit the spectral index
$\gamma = -2.86 \pm .03$ using all spatial frequencies $ k < 0.07 $
arcmin$^{-1}$.

It is obvious from Fig. \ref{Fig_spec_aver} (red) that uncertainties in
the derived power spectrum increase rapidly for $ k \ga 0.08 $
arcmin$^{-1}$. This limit depends on the signal-to-noise ratio of the
observations but reflects also the limited spatial sensitivity of the
100-m telescope. For our analysis in the 3C~196 field we strictly use in
the following only data at spatial frequencies $ k < 0.07 $
arcmin$^{-1}$. This limit is indicated in all plots with a horizontal
line.

\begin{figure}[tbp]
   \centering
   \includegraphics[width=6.5cm,angle=-90]{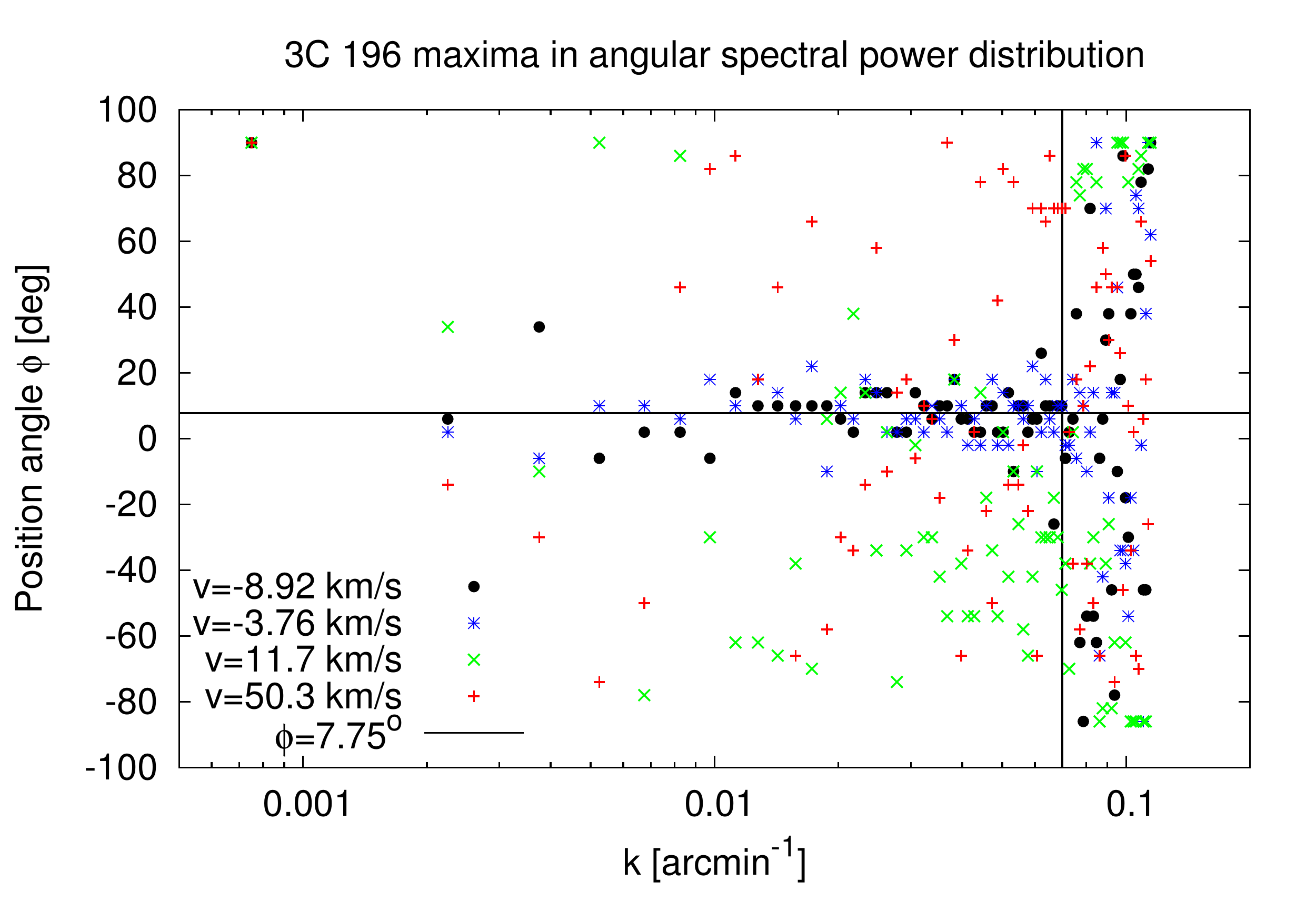}
   \caption{Angular peak power distribution for single channels at
     velocities of $v_{\rm LSR} = -8.92, -3.76, 11.69$, and 50.3
     \kms. Data for $ k > 0.07$ arcmin$^{-1}$ (vertical line) are
     affected by the instrumental noise. The average position angle for
     $v_{\rm LSR} = -8.92$ and -3.76 \kms is $\phi = 7\fdg5$.  }
   \label{Fig_spec_angle}
\end{figure}

\begin{figure}[htb]
   \centering
   \includegraphics[width=6.5cm,angle=-90]{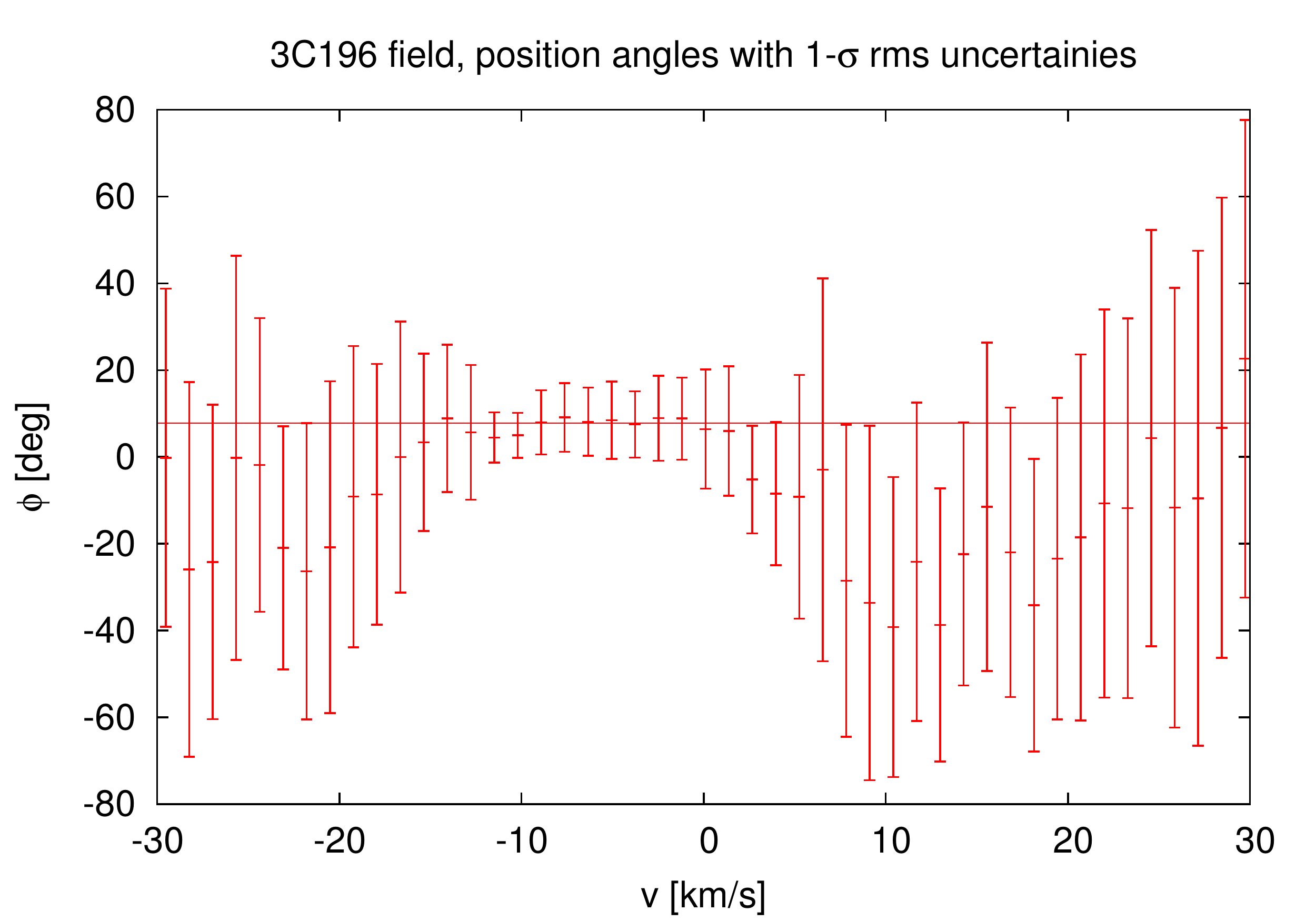}
   \caption{Position angles fit for each channel at
     spatial frequencies $ 0.005 < k < 0.05 $. The average position
     angle $\phi = 7\fdg5$ from Fig. \ref{Fig_spec_angle} is indicated
     with a horizontal line. }
   \label{Fig_3C196_angle}
\end{figure}

\subsection{Angular power distribution}
\label{Angular}

To determine anisotropies, we first search for an average position
angle. We average data within sectors $\Phi \pm \delta\Phi$ to obtain
the power $P(\Phi,k)$ at position angles $\Phi$. Some care is necessary
here, since we need enough samples per sector for statistical
significant results. Position angles are ambiguous by 180\degr. Dividing
this range into $n$ sectors implies that the noise increases by a factor
of $\sqrt n$. Searching for an average position angle over a range of
spatial frequencies, this amplification is balanced by averaging over
$m$ spatial frequencies, the formal increase of the uncertainties is
then $\sqrt{n/m}$.  After some tests we have chosen $n = 45$, since the
typical number of samples in spatial frequencies turned out to be $m
\sim 23$, hence $\sqrt{n/m} \sim \sqrt 2$ with our choice of
$2\delta\Phi = 4\degr$. Thus the selection of $\delta\Phi$ is not
critical as long as position angles are correlated over a certain range
in spatial frequencies.

Figure \ref{Fig_spec_angle} displays some characteristic results of our
search for anisotropies. This plot displays the position angle $\Phi$
for the peak power emission. For two channels at velocities $v_{\rm LSR}
= -8.92$ and -3.76 \kmss we find a rather narrow and well defined
distribution for the peaks around an average of $\Phi = 7\fdg5$ for
spatial frequencies between $0.005 < k < 0.05 $ arcmin$^{-1}$ (black and
blue points). The channel map at $v_{\rm LSR} = 11.69$ \kmss (green) has
quite a different angular power distribution with an average at $\Phi
\sim -45 \degr$, but with a large scatter. We also display the
distribution for a $v_{\rm LSR} = 50$ \kmss (red), this case shows a
random scatter. The distribution at high spatial frequencies, $ k > 0.07
$ arcmin$^{-1}$ is in general very noisy and the first data point at $ k
= 0.00075 $ arcmin$^{-1}$ is affected by smoothing from the apodization
function. 

We analyse henceforth data in the range $ 0.005 < k < 0.05 $
arcmin$^{-1}$ for an automated determination of anisotropies in the
spectral power distribution and fit average position angles for all
velocity channels. Figure \ref{Fig_3C196_angle} shows for velocities
$-11.5 < v_{\rm LSR} < -1.2$ \kmss a constant position angle with a
typical dispersion of 8 \kms and a weighted mean of $\Phi =
6\fdg9$. This angle corresponds to the position angle $\Phi_{\rm mag}
\sim 97\degr$ of the \hi filaments, shown in Fig.  \ref{Fig_cubep_USM}
and is aligned with the magnetic field \citep{Zaroubi2015}. Thus we use
in the Fourier plane $\Phi_{\perp} = 7\fdg5$.

We can draw here our first conclusion about the nature of the observed
MHD turbulence. According to \citet{Kandel2016} Alfv{\'e}n and slow
modes cause the eddies to be elongated in direction parallel to the
sky-projected mean magnetic field, while for fast modes the elongation
is perpendicular. We find elongation parallel to the magnetic field,
hence no evidence for fast modes.

\subsection{Anisotropic power spectra }
\label{Anisotropic}

We verified that the peaks in the angular power distribution are
associated with minima in the power distribution at angles offset by
typically 90\degr. We define for each velocity channel $v_{\rm LSR}$ the
power anisotropy
\begin{eqnarray}\nonumber
Q(k,v_{\rm LSR}) = P(\Phi_{\perp},k,v_{\rm LSR}) / P(\Phi_{\parallel},k,v_{\rm LSR}) \\
= P_{\perp}(k,v_{\rm LSR})/P_{\parallel}(k,v_{\rm LSR}),
\label{eq:Q}
\end{eqnarray}
such that $Q = 1$ denotes isotropy, while larger values of $Q$ mean larger
anisotropy with eddies elongated parallel to the magnetic field.  We
calculate $Q$ by integrating in sectors $ 0\degr < \Phi_{\perp} <
15\degr $ and $ 90\degr < \Phi_{\parallel} < 105\degr $. 

The anisotropies $Q(k,v_{\rm LSR})$ show quite some scatter. We
define therefore a geometric mean anisotropy 
\begin{equation}
  Q_{\rm aver}(v_{\rm LSR}) = {\rm exp} \left[ \frac{1}{n} 
\sum_{i=1}^n {\rm ln}~ Q(k_i,v_{\rm LSR}) \right],
\label{eq:Qav}
\end{equation}
for $n$ data-points in the range $ 0.005 < k_i < 0.05 $ arcmin$^{-1}$.
$Q_{\rm aver}(v_{\rm LSR})$ is close to the median anisotropy $Q_{\rm
  median}(v_{\rm LSR})$ for this spatial frequency range.

\begin{figure}[tbp]
   \centering
   \includegraphics[width=6.5cm,angle=-90]{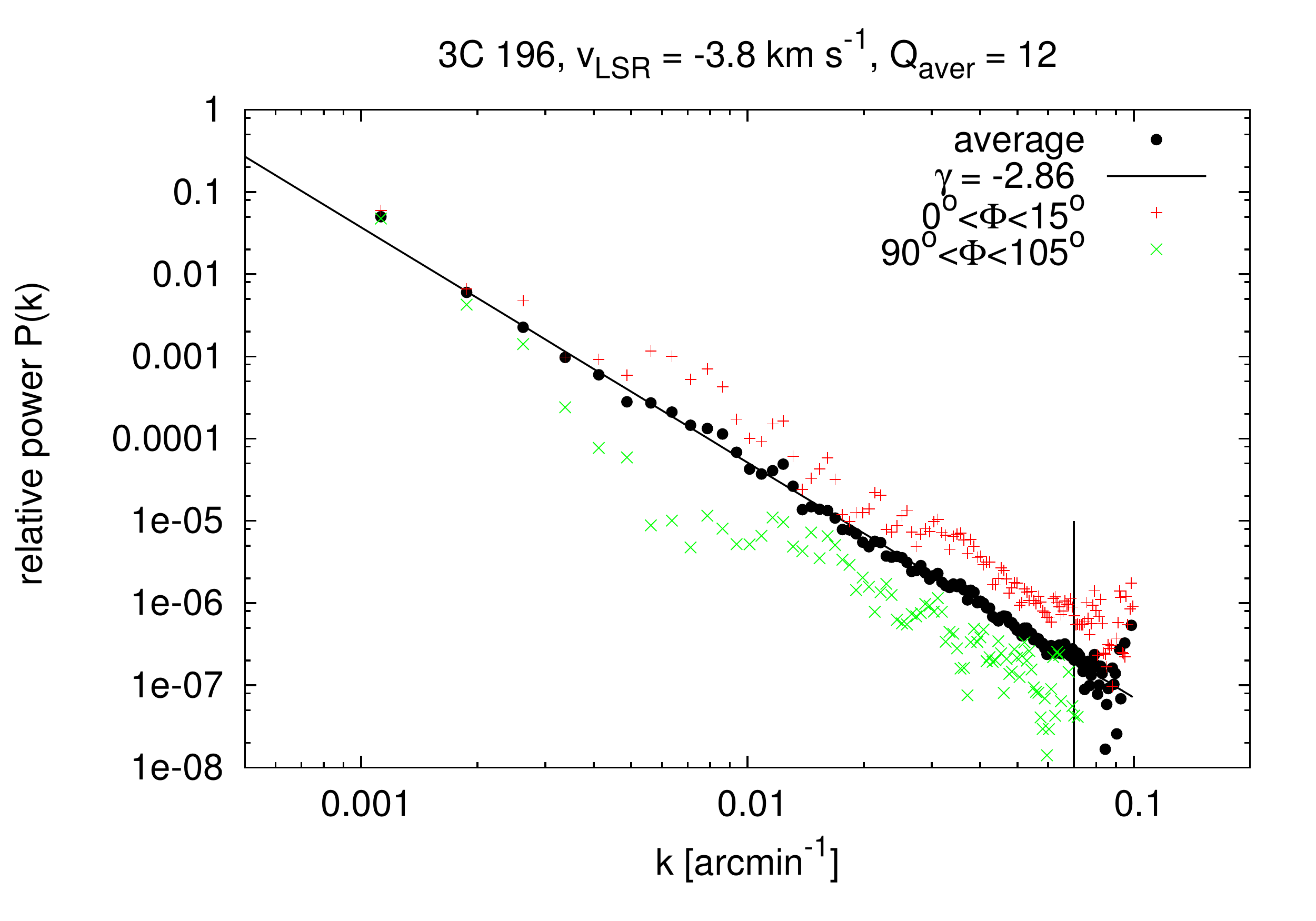}
   \caption{Average power spectrum observed for $v_{\rm LSR} = -3.76$
     \kmss and fit power law (black) with $\gamma = -2.86 \pm
     0.03 $ for $ k < 0.07$ arcmin$^{-1}$ (vertical line). In addition
     the power spectrum for $0\degr < \Phi < 15\degr $ (red) and
     $90\degr < \Phi < 105\degr $ (green) is given. The average
     anisotropy factor is $Q_{\rm aver} = 12.1$ }
   \label{Fig_spec_37}
\end{figure}

\begin{figure}[tbp]
   \centering
   \includegraphics[width=6.5cm,angle=-90]{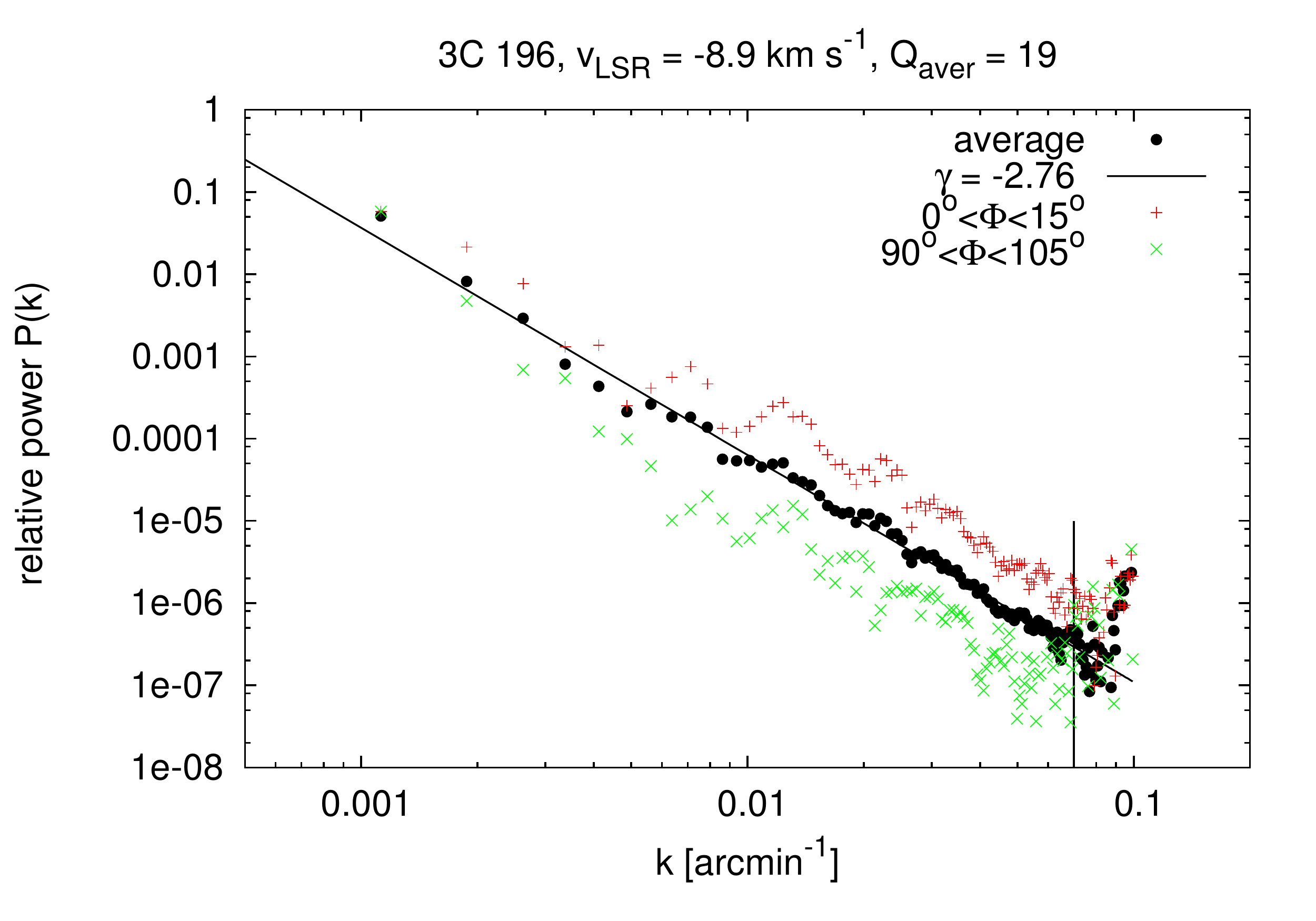}
   \caption{Average power spectrum observed for $v_{\rm LSR} = -8.92$
     \kmss (black dots) and fit power law with $\gamma = -2.76 \pm
     0.04 $ for $ k < 0.07$ arcmin$^{-1}$ (vertical line). In addition
     the power spectrum for $0\degr < \Phi < 15\degr $ (red) and
     $90\degr < \Phi < 105\degr $ (green) is given. The average
     anisotropy factor is $Q_{\rm aver} = 18.8$ }
   \label{Fig_spec_33}
\end{figure}

\begin{figure}[tbp]
   \centering
   \includegraphics[width=6.5cm,angle=-90]{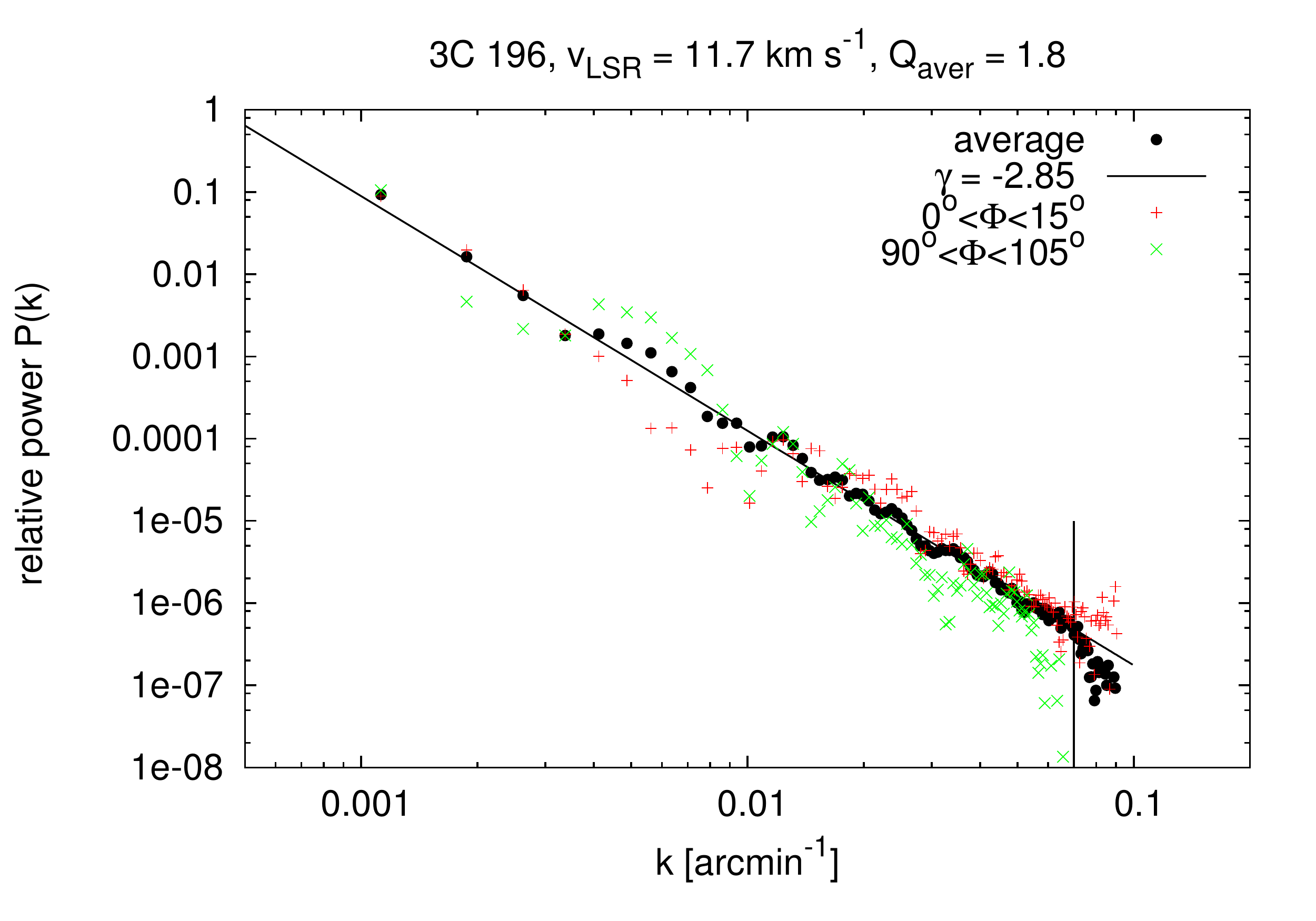}
   \caption{Average power spectrum observed for $v_{\rm LSR} = 11.69$
     \kmss (black dots) and fit power law with $\gamma = -2.85\pm
     0.03 $ for $ k < 0.07$ arcmin$^{-1}$ (vertical line). In addition
     the power spectrum for $0\degr < \Phi < 15\degr $ (red) and
     $90\degr < \Phi < 105\degr $ (green) is given. The average
     anisotropy factor is $Q_{\rm aver} = 1.8$. }
   \label{Fig_spec_49}
\end{figure}

Figures \ref{Fig_spec_37} and \ref{Fig_spec_33} show power spectra at
$v_{\rm LSR} = -3.76$ \kmss and $v_{\rm LSR} = -8.92$ \kmss with rather
strong average anisotropies, $Q_{\rm aver} = 12.1$ and 18.8.  Particular
strong local anisotropies are visible in Fig. \ref{Fig_spec_37}. For
comparison we add in Fig. \ref{Fig_spec_49} a plot for $v_{\rm LSR} =
11.69$ \kms, a case without significant anisotropy ($Q_{\rm
  aver} = 1.8 $), see also Fig. \ref{Fig_spec_angle} (green).

To summarize our finding from Figs. \ref{Fig_spec_37} to
\ref{Fig_spec_49}, we find significant anisotropies at channels with
velocities around $v_{\rm LSR} \sim -4$ \kmss in contrast to $v_{\rm LSR}
\sim 11$ \kms, where no significant anisotropies are detected. Except
for some deviations in the range $ 0.005 \la k \la 0.02 $ arcmin$^{-1}$,
the red and green power spectra appear in logarithmic presentation just
to be displaced from the isotropic distribution, sharing a similar slope
but on average with slight inclinations in opposite directions.

\subsection{\hi parameter dependencies on  $v_{\rm LSR}$}
\label{vlsr}

\begin{figure}[tbp]
   \centering
   \includegraphics[width=6.5cm,angle=-90]{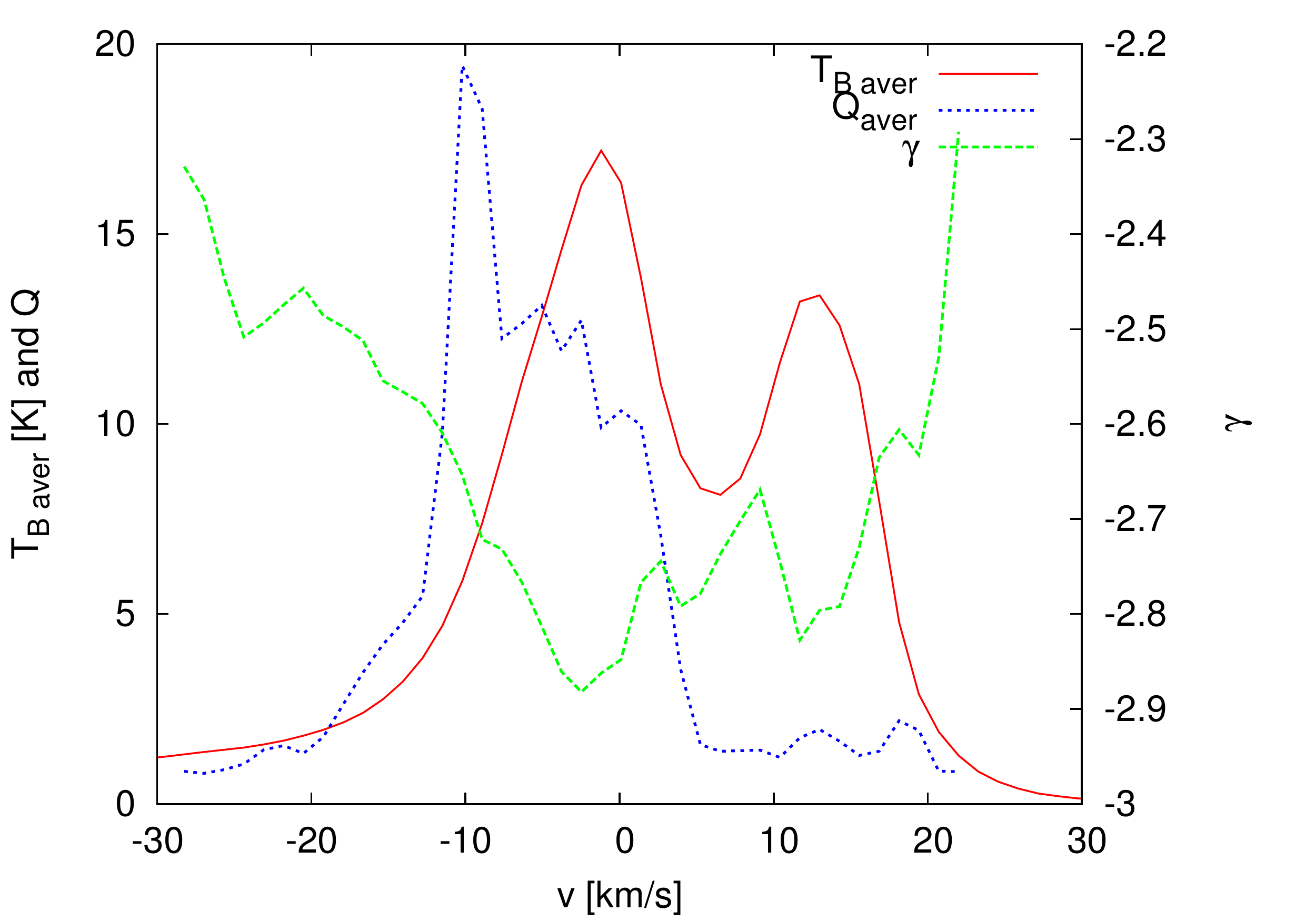}
   \caption{Comparison between average brightness temperature $T_{\rm
       B\, aver}$ (red), anisotropy factor $Q_{\rm aver}$ (blue), and
     spectral index $\gamma$ (green) for the 3C~196 field.  }
   \label{Fig_overview}
\end{figure}

Figure \ref{Fig_overview} summarizes global properties of the \hi gas
distribution in the 3C~196 field. To characterize the \hi emission
  we calculate the average brightness temperature as the weighted mean
  over the apodized brightness temperature, 
\begin{equation}
T_{\rm B\, aver} =  \frac{\sum T_{\rm B} \cdot W} { \sum W}. 
\label{eq:Taver}
\end{equation}
Here $W$ is the apodization weight due to the 50\% Tukey taper function.
The \hi distribution (plotted in red) shows two major components in
emission, peaking at $v_{\rm LSR} = -1.2 $ \kmss and at $v_{\rm LSR} =
13$ \kms. For comparison we plot also the derived anisotropy $Q_{\rm
  aver}(v_{\rm LSR})$ (blue) and the average spectral index
$\gamma(v_{\rm LSR})$ (green). Anisotropies are strongest for channels
with best defined position angles (compare Fig. \ref{Fig_3C196_angle}
with Fig. \ref{Fig_overview}). The major local minima in the spectral
index distribution $\gamma$ coincide with the peaks of the \hi
components. A similar trend was found by \citet{Stanimirovic1999} in the
SMC.

The velocities of CNM components tend to be closely related to the
center velocities of the warm neutral medium (WNM)
\citep{Kalberla2016}. It is also well established that the velocity
dispersion of the WNM is significantly larger than the dispersion of the
CNM components \citep[e.g.,][]{Heiles2003}.  The CNM is clumpy while the
WNM has a diffuse structure. Thus the CNM adds extra power at high
spatial frequencies and for a mix of CNM and WNM gas one expects that
the spectral index should be flattest close to the center velocity of
the WNM component where the CNM is most pronounced. Correspondingly
\citet{Martin2015}, after decomposing WNM and CNM components, find for
the column density distribution that the spectral index of the CNM is
shallower than that of the WNM.

\begin{figure}[tbp]
   \centering
   \includegraphics[width=6.5cm,angle=-90]{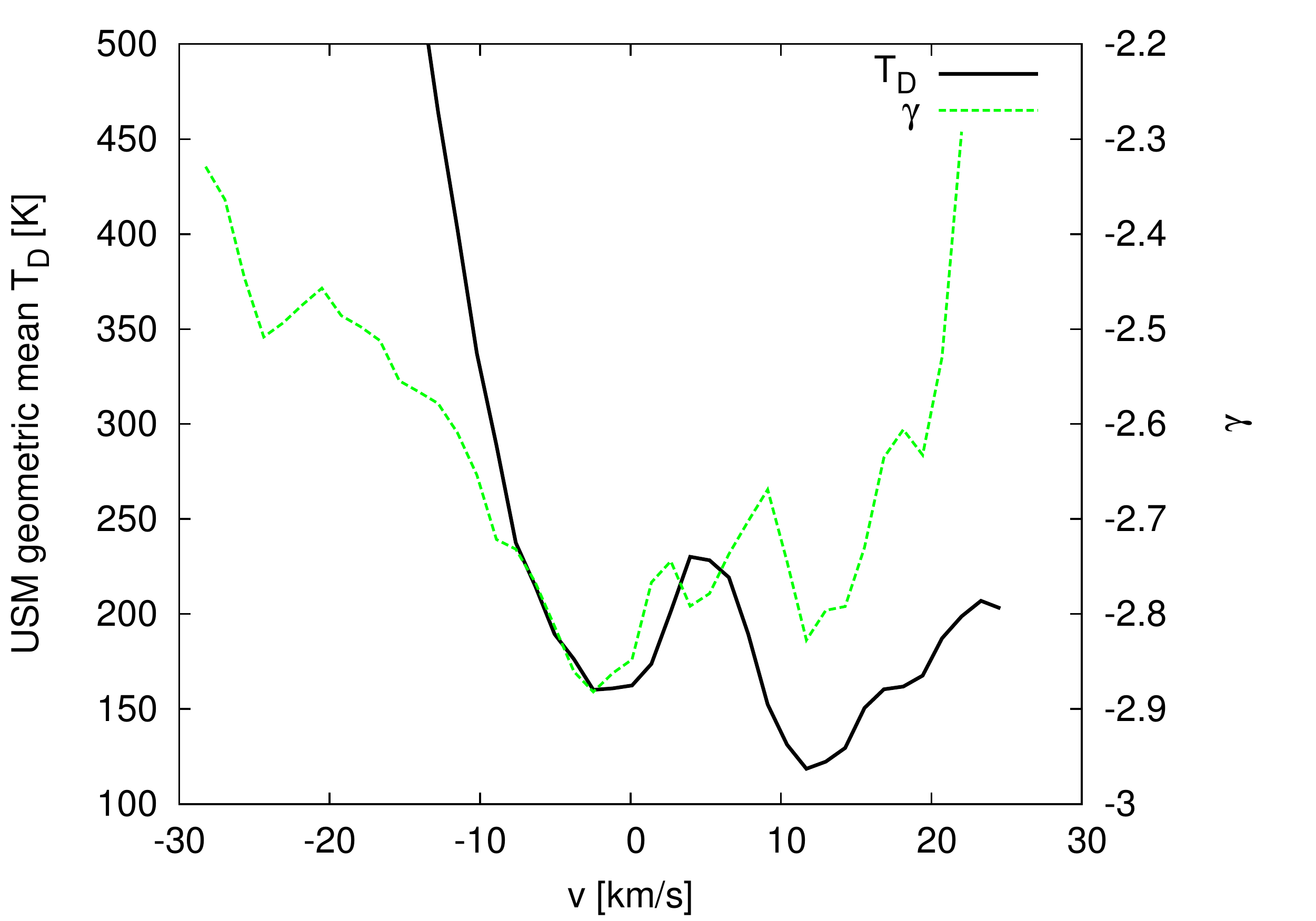}
   \caption{Comparison between the geometric mean Doppler temperature
     $T_{\rm D}$ (black) and the spectral index $\gamma$ (green dashed)
     for the 3C~196 field.  }
   \label{Fig_overview2}
\end{figure}

For the observed mix of WNM and CNM we find the opposite. Consistent
with \citet{Stanimirovic1999}, the spectral index for thin slices in
velocity is steepest at the center velocity of the WNM component.
Figure \ref{Fig_overview2} shows that the spectral index steepens as the
CNM Doppler temperature (hence the velocity dispersion) decreases.  We
find no indications for a break in the spectral power distribution that
could be attributed to either the CNM or the WNM. This suggests that
turbulence in the WNM and CNM cannot be considered
independently. Turbulence may induce the formation of CNM when the WNM
is pressurized and put in a thermally unstable state
\citep{Saury2014}. Since turbulence is probably playing a major r\^ole in
regulating the ratio between WNM and CNM, phase transitions must be the
key in understanding the steepening of the spectral index. The usual
assumption is that steep spectra correspond to random fields dominated by
large-scale fluctuations \citep{Lazarian2000}. In our case the \hi
filaments, visible over many degrees, may be indicative for an event
that could have driven phase transitions.  

The anisotropy measure $Q_{\rm aver}$, does not correlate with
$T_{\rm B\, aver}$, nor with $\gamma$.
\citet{Kalberla2016} found that the CNM is mostly organized in cold \hi
filaments with Doppler temperatures $T_{\rm D} \sim 223 $ K. To check,
whether $T_{\rm D}$ might be correlated with $\gamma$ or $Q_{\rm aver}$, we
determined the geometric mean $T_{\rm D}$ for the CNM filaments toward the
3C~196 field. Again we find no correlation with $Q_{\rm aver}$, but
Fig. \ref{Fig_overview2} shows that for the coldest parts of the filaments
also the spectral index $\gamma$ (green) is steep.

\hi absorption line data against the continuum source 3C~196 show two
components, at a velocity of $ v_{\rm LSR} = -2.2 $ \kms with an optical
depth of $\tau = 0.086$ and at $ v_{\rm LSR} = 15.3 $ \kms~ with $\tau =
0.130$. The component at $ v_{\rm LSR} = 15.3 $ \kms~ has a spin
temperature of 80.9 K \citep{Mebold1982}. This temperature compares well
with the low geometric mean $T_{\rm D}$ determined by us for the whole
field, but we note that we find the lowest geometric mean $T_{\rm D}$ at $
v_{\rm LSR} = 11.69 $ \kms.  \citet{Crovisier}, using the Nan\c{c}ay
telescope, find evidence for absorption at $ v_{\rm LSR} = -7.0, -2.2 $,
and 15.7 \kms. The additional component at $ v_{\rm LSR} = -7.0$ \kmss
most probably is due the fan beam of the Nan\c{c}ay telescope, extending in
north/south direction and thus being more sensitive to filaments
oriented in the same direction.

\subsection{Thick slices, velocity centroid}
\label{Thick}

So far we analyzed only individual channel maps, in each case covering a
range of $\delta v_{\rm LSR} = 1.44 $ \kmss in velocity, the limiting
resolution of the EBHIS \citep{Winkel2016a}. This channel width formally
corresponds to a Doppler temperature of $T_{\rm D} \sim 50 $ K. The
median Doppler temperature of the CNM is $T_{\rm D} \sim 223 $ K
\citep{Kalberla2016}. Figure \ref{Fig_overview2} discloses that some of
the filaments in the 3C~196 field are colder ($T_{\rm D} \ga 160$) than
the median. 

A velocity channel $\delta v_{\rm LSR} = 1.44 $ \kmss is, according to
the definition of \citep{Lazarian2000}, thin if the line-of-sight
FWHM velocity width of the gas is larger than the channel
width. Otherwise the observed slice is thick. Our observational setup
implies that an individual velocity channel is thin for most of the gas
but in the limit it may be thick for the coldest CNM components. 

\begin{figure}[htp]
   \centering
   \includegraphics[width=6.5cm,angle=-90]{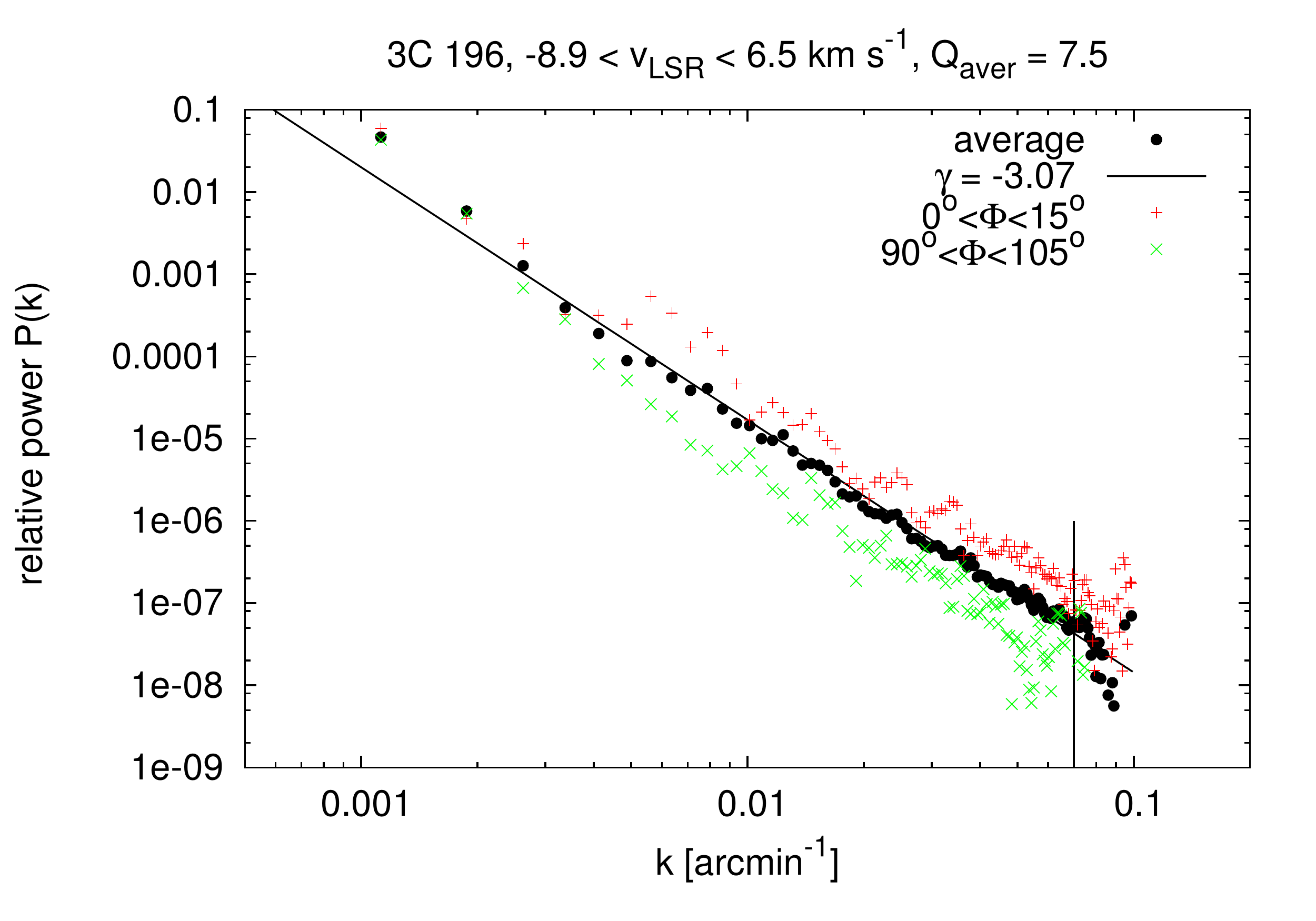}
   \caption{Average power spectrum for a thick slice, $-8.9 < v_{\rm
       LSR} < 6.5$ \kmss (black dots) and fit power law with $\gamma
     = -3.07 \pm 0.04 $ for $ k < 0.07$ arcmin$^{-1}$ (vertical
     line). In addition the power spectrum for $0\degr < \Phi < 15\degr
     $ (red) and $90\degr < \Phi < 105\degr $ (green) is given. The
     average anisotropy factor is $Q_{\rm aver} = 7.5$ }
   \label{Fig_spec_33_45}
\end{figure}

\begin{figure}[htb]
   \centering
   \includegraphics[width=6.5cm,angle=-90]{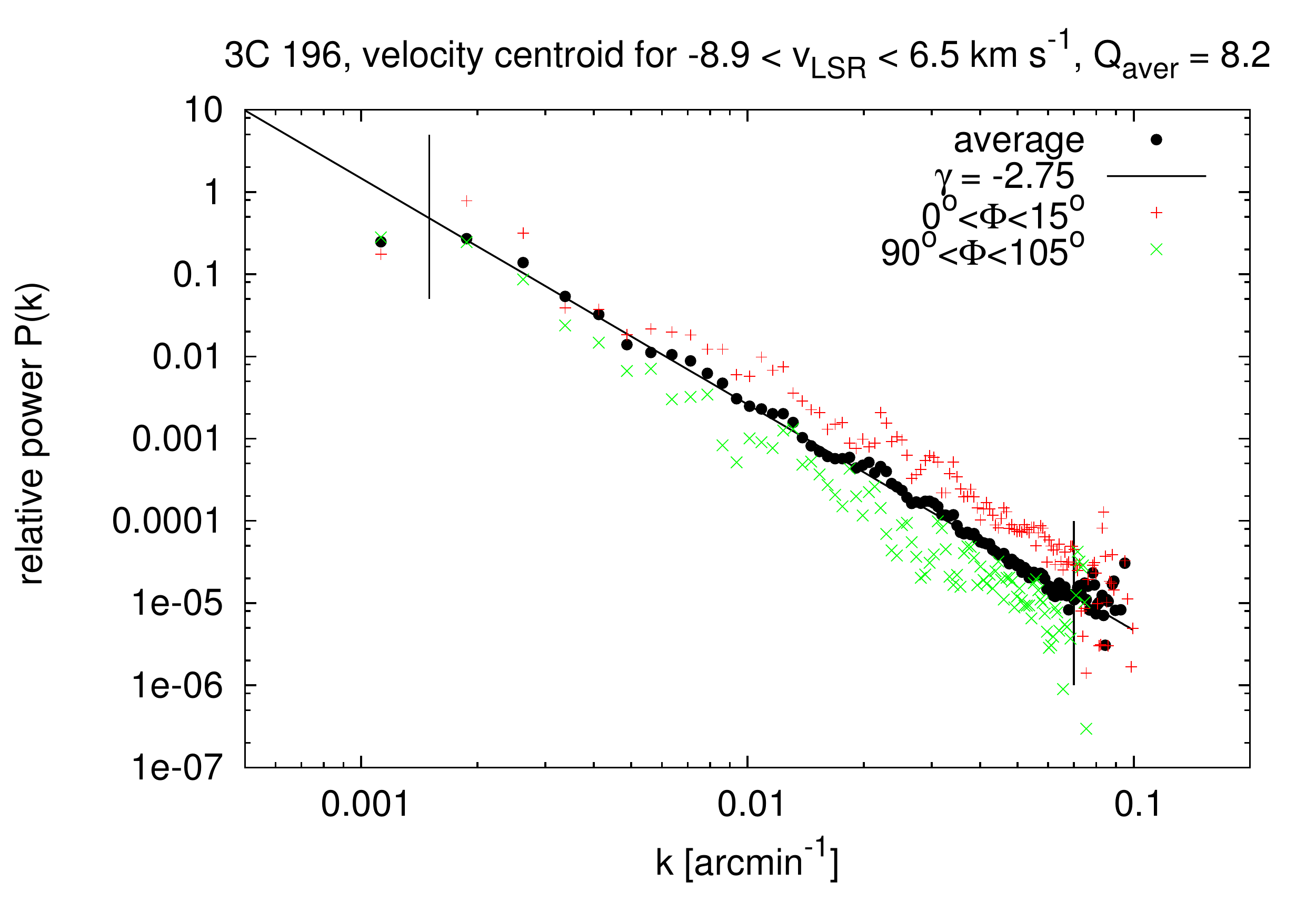}
   \caption{Average power spectrum for the velocity centroid in the
     range $-8.9 < v_{\rm LSR} < 6.5$ \kmss (black dots) and fit
     power law with $\gamma = -2.75 \pm 0.04 $ for $ 0.0015< k < 0.07$
     arcmin$^{-1}$ (vertical lines). In addition the
     power spectrum for $0\degr < \Phi < 15\degr $ (red) and $90\degr <
     \Phi < 105\degr $ (green) is given. The average anisotropy factor
     is $Q_{\rm aver} = 8.2$. }
   \label{Fig_spec_33_45_centroid}
\end{figure}

Filamentary features are mostly sheets, seen tangentially, with a median
thickness between 0.1 and 0.3 pc
\citep{Heiles2005,Kalberla2016}. Because the sheets are cold, we observe
along the line of sight different radial velocities which causes in many
cases an apparent positional shift perpendicular to the filament
\citep[see Figs. 10, 25, and 26 of][]{Kalberla2016}. The 3C~196 field
under investigation contains several filaments (or sheets),
approximately oriented parallel to each other.

The average \hi emission contains two components. To cover the major
part of the anisotropic filamentary range, we integrate over the
channels for $ -8.9 < v_{\rm LSR} < 6.5 $ \kmss (see
Fig. \ref{Fig_overview}). This slice is thick \citep{Lazarian2000} with
respect to the CNM but, because of blending with the isotropic component
at $v_{\rm LSR} \sim 11$ \kms, does not cover the full width of the WNM
as requested by \citet{Lazarian2000} for a representative thick slice. A
general problem with the \hi distribution is that several \hi components
may be superposed along the line of sight. An overlap in velocity may
then, as in our case, smear out or merge individual \hi features. In
this case it is difficult to derive a representative thick slice. Figure
\ref{Fig_spec_33_45} shows a power law $\gamma = -3.07 \pm 0.04 $ with
$Q_{\rm aver} = 7.5$.  For the velocity centroid over the same velocity
range we determine $\gamma = -2.75 \pm 0.04 $, see Fig.
\ref{Fig_spec_33_45_centroid}. The anisotropy is this case slightly
larger, $Q_{\rm aver} = 8.2$.

According to \citet[][Eq. 28]{Lazarian2000}, thin and thick sheets
should have distinct different power indices. We find only a moderate
steepening with slice thickness by $\delta \gamma \sim 0.2$, comparable to
\citet{Stanimirovic2001}. Anisotropies are noticeable but weaker than for
most of the prominent thin slices. Velocity centroids, proposed by
\citet{Esquivel2005}, are apparently not always the best choice when
studying anisotropies.

\begin{figure}[tbp]
   \centering
   \includegraphics[width=6.5cm,angle=-90]{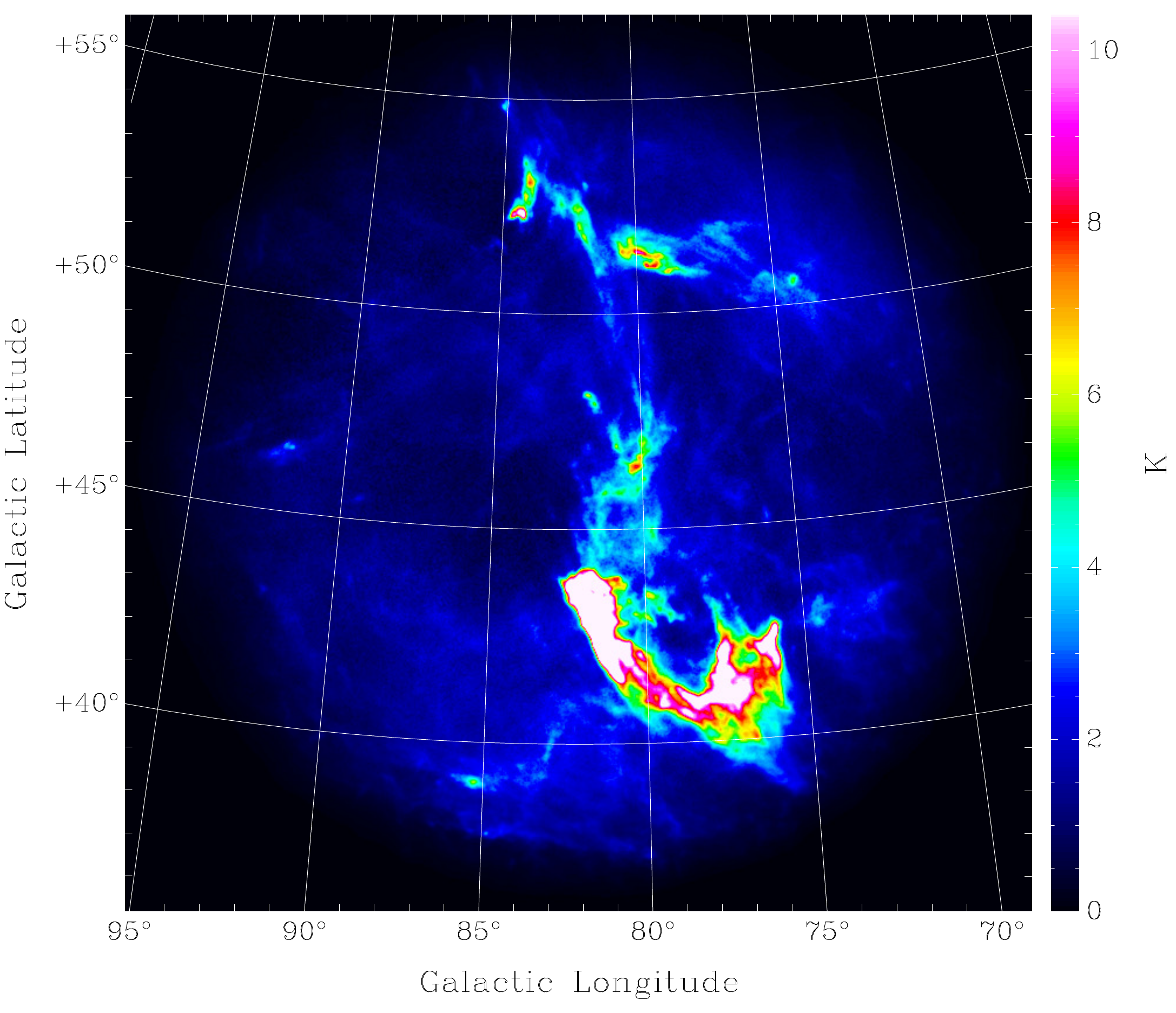}
   \caption{Apodized \hi distribution in direction to the FN1 field at a
     velocity of $v_{\rm LSR} = 3.97$ \kms. This map is used to test the
     methodology of our approach in case of diffuse HI structures.}
   \label{Fig_FN1_Image}
\end{figure}

\section{Diffuse HI structures at high Galactic latitudes}
\label{Diffuse}

The 3C~196 spectral component at $v_{\rm LSR} \sim 11$ \kmss shows in
Fig. \ref{Fig_spec_49} only weak anisotropies with $Q_{\rm aver} =
1.8$. To get an estimate, how far power spectra and anisotropies in
direction to the 3C~196 field deviate from typical diffuse \hi structures,
we decided to analyze for comparison a field without obvious filaments
or anisotropies.

\citet{Miville-Deschenes2007} studied such a case, a very diffuse \hi
filament located in the extra-galactic window known as Firback North 1
\citep[FN1,][]{Dole2001} at high Galactic latitudes ($l = 84\degr, b =
45\fdg1 $).  They used data from the Leiden/Dwingeloo Survey
\citep[LDS,][]{LDS} and from the Green Bank Telescope (GBT) and
  calculated the integrated emission and the velocity centroid for
  $-86.2 < v_{\rm LSR} < 47.8$ \kms. We are going to use this velocity
  range, representing a thick slice, for reference.

\begin{figure}[hbp]
   \centering
   \includegraphics[width=6.5cm,angle=-90]{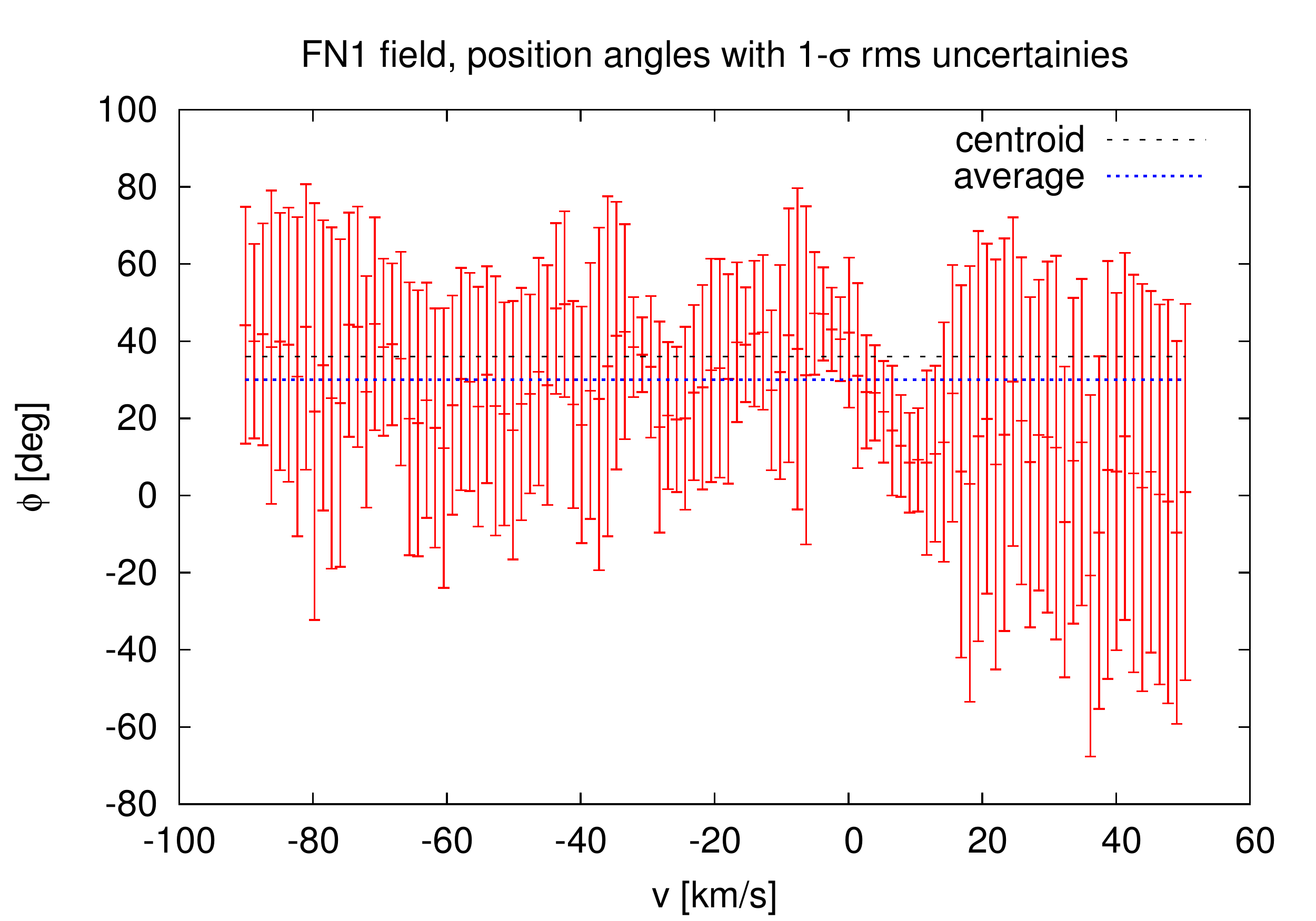}
   \caption{Position angles fit for each channel at spatial
       frequencies $ 0.005 < k < 0.05 $. The position angles for the
       centroid and the average emission, $ \phi = 36\degr \pm 19\degr $
       and $ \phi = 30\degr \pm 15\degr$, respectively, are also
       indicated. } 
   \label{Fig_FN1_angle}
\end{figure}

\begin{figure}[tbp]
   \centering
   \includegraphics[width=6.5cm,angle=-90]{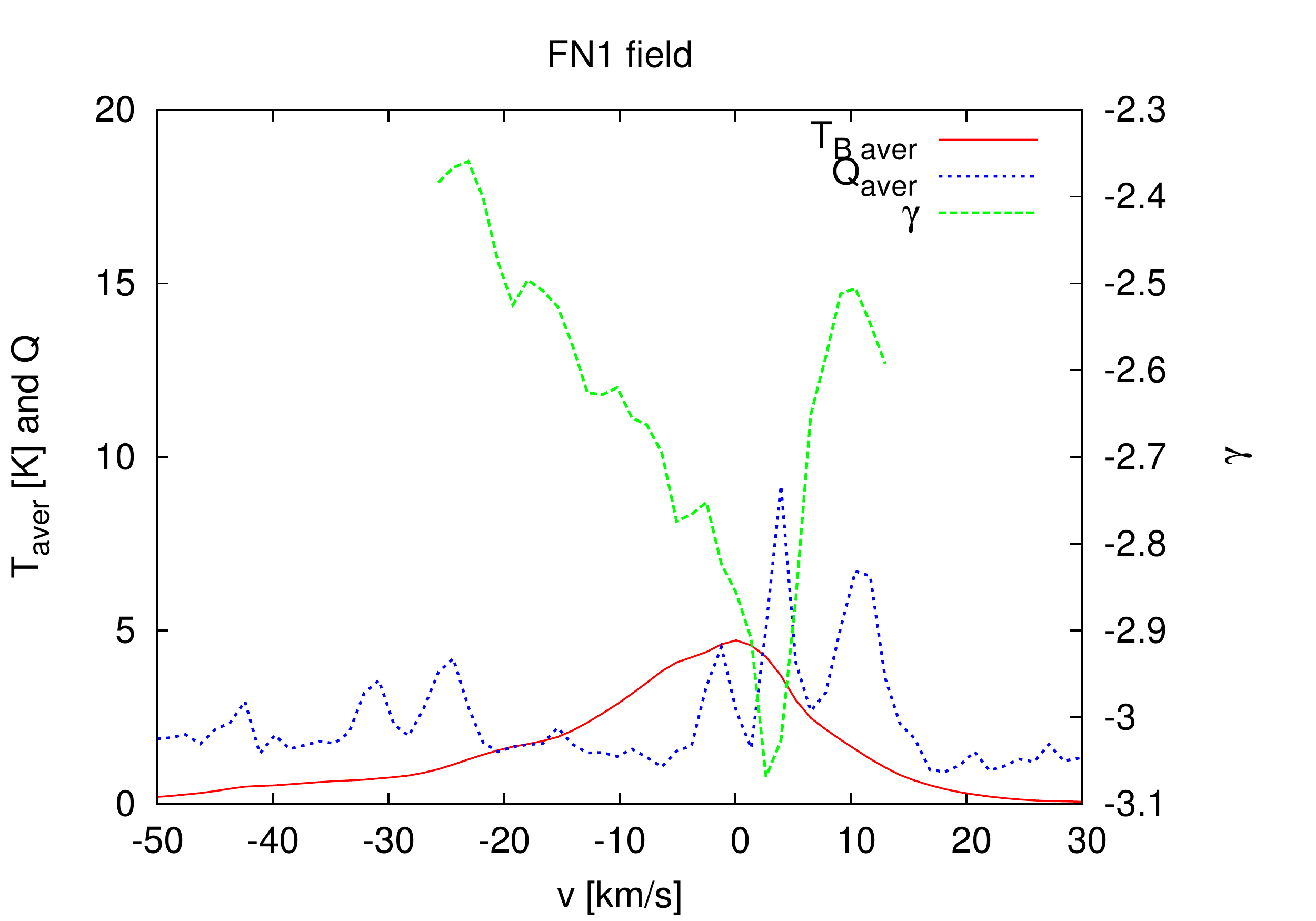}
   \caption{Comparison between average brightness temperature $T_{\rm
       B\, aver}$ (red), anisotropy factor $Q_{\rm aver}$ (blue), and
     spectral index $\gamma$ (green) for the FN1 field.  }
   \label{Fig_FN1_overview}
\end{figure}

For the FN1 field we applied the same data-processing pipeline as
described in Sect. \ref{data_analysis}. We used the EBHIS to generate a
field with 38\degr in diameter and apodized this with an inner diameter
of 19\degr, see Fig. \ref{Fig_FN1_Image}. We derived the angular
  power distribution and determined position angles for all emission
  channels. Figure \ref{Fig_FN1_angle} shows angles $ 45\degr \ga \phi
  \ga 0\degr$, but typically with a huge dispersion. The velocity range
  $ -5 \la v_{\rm LSR} \la 10 $ \kmss is outstanding, here we find better
  defined position angles, but with a systematical rotation by almost
  40\degr. The position angles for the centroid and the average emission
  are $ \phi = 36\degr \pm 19\degr $ and $ \phi = 30\degr \pm 15\degr$,
  respectively, also indicated in Fig. \ref{Fig_FN1_angle}.

  Figure \ref{Fig_FN1_overview} displays the average emission (red)
  according to Eq. \ref{eq:Taver}, average anisotropies (blue) and the
  fit average spectral index $\gamma$ (green).  The emission is weak
  with a long extended tail to negative velocities; we plot only the
  most significant range $ -50 \la v_{\rm LSR} \la 30 $ \kms. To avoid
  spurious results, we determined spectral indices $\gamma$ only for
  $T_{\rm B\, aver} > 1 $ K.

\begin{figure}[tbp]
   \centering
   \includegraphics[width=6.5cm,angle=-90]{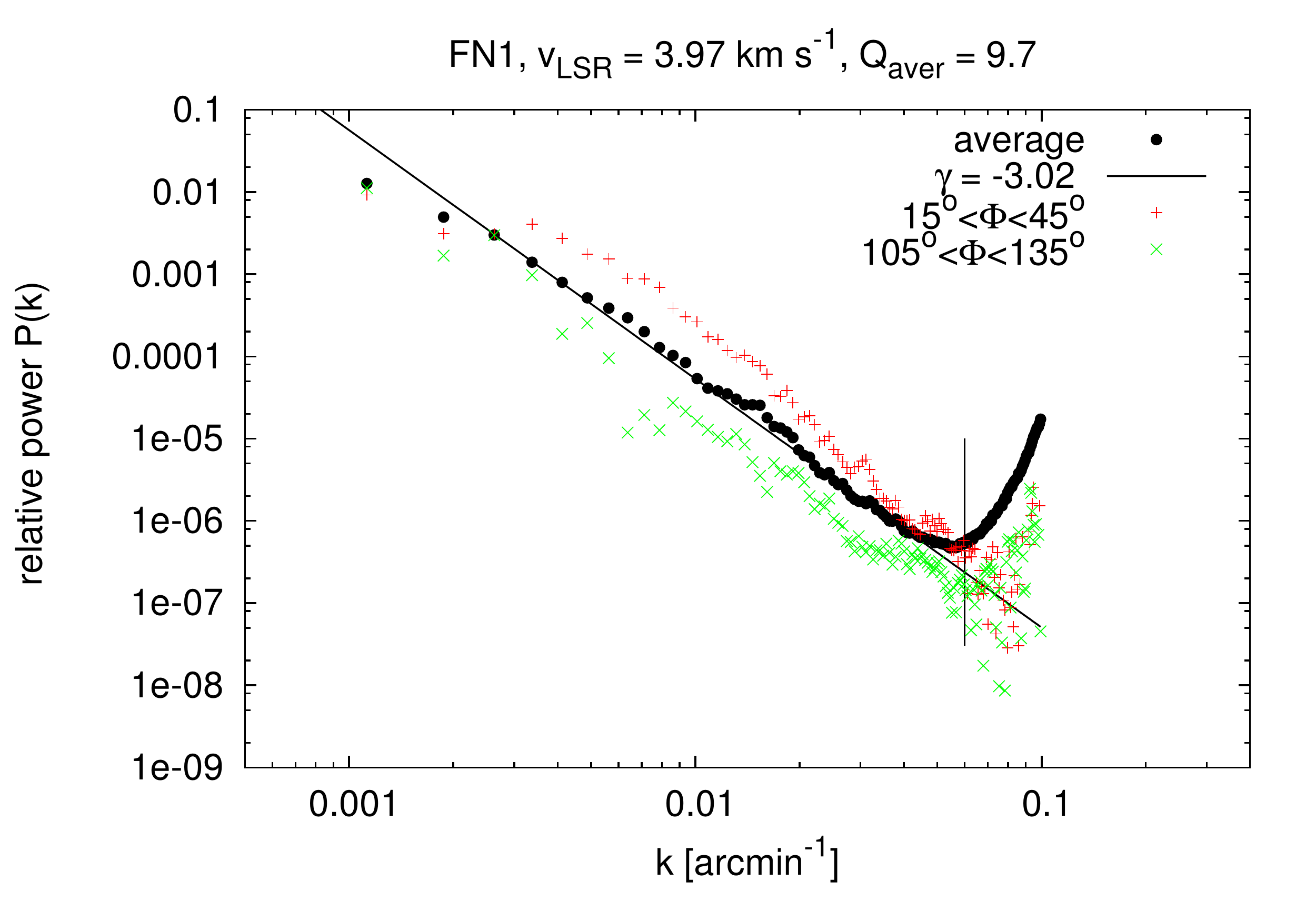}
   \caption{Average power spectrum at $v_{\rm LSR} = 3.97$ \kmss and
     fit power law (black) with $\gamma = -3.02\pm 0.03 $ for $ k < 0.06$
     arcmin$^{-1}$ (vertical line). In addition the power spectrum for
     $15\degr < \Phi < 45\degr $ (red) and $115\degr < \Phi < 135\degr $
     (green) is given. The average anisotropy factor is $Q_{\rm aver} =
     9.7$. }
   \label{Fig_FN1_74}
\end{figure}

\begin{figure}[tbp]
   \centering
   \includegraphics[width=6.5cm,angle=-90]{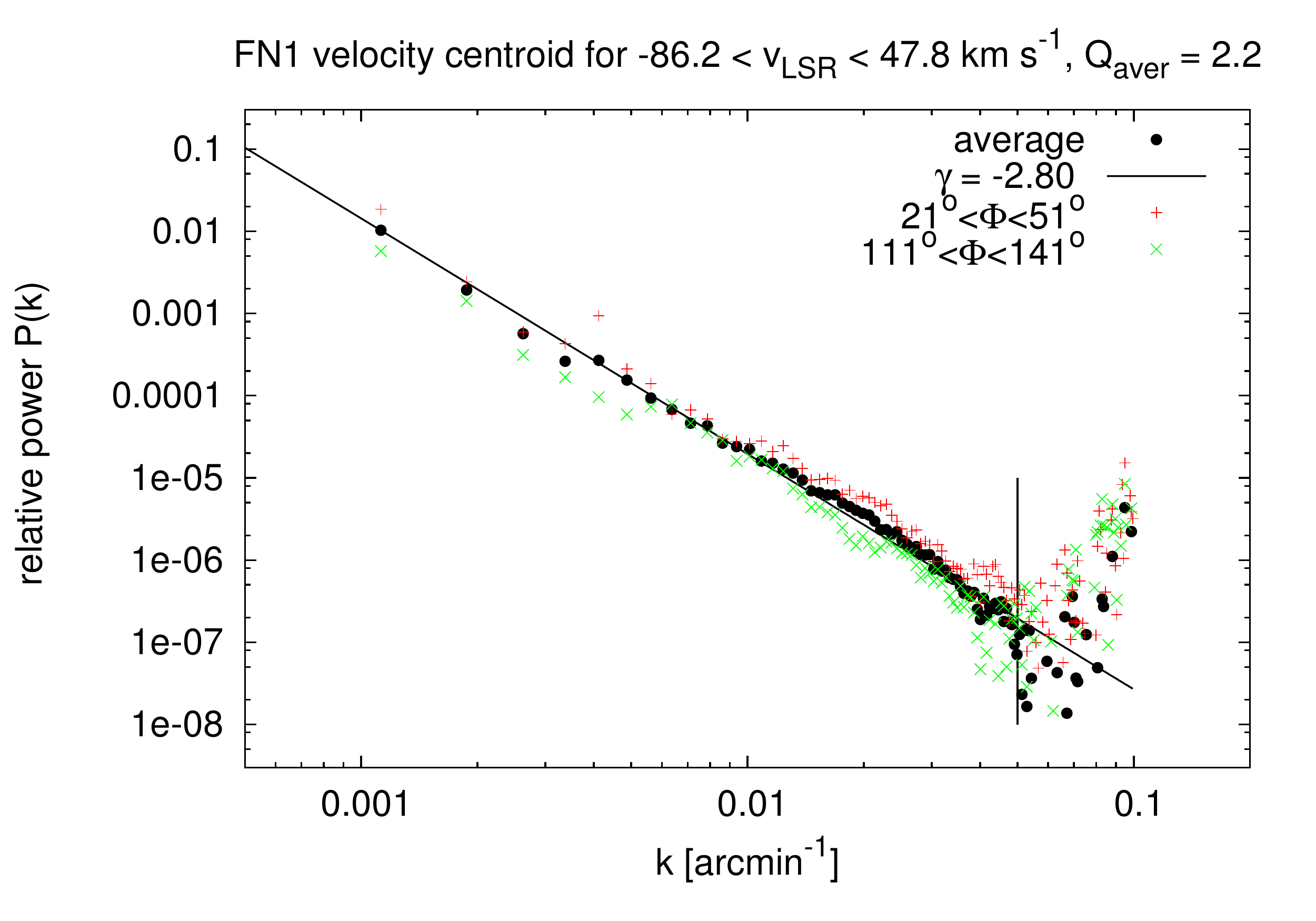}
   \caption{Average power spectrum for the velocity centroid and fit
     power law (black) with $\gamma = -2.86\pm 0.06 $ for $ k < 0.05$
     arcmin$^{-1}$ (vertical line). In addition the power spectrum for
     $30\degr < \Phi < 60\degr $ (red) and $120\degr < \Phi < 150\degr $
     (green) is given. The average anisotropy factor is $Q_{\rm aver} =
     2.0$. }
   \label{Fig_FN1_centroid}
\end{figure}

\begin{figure}[tbp]
   \centering
   \includegraphics[width=6.5cm,angle=-90]{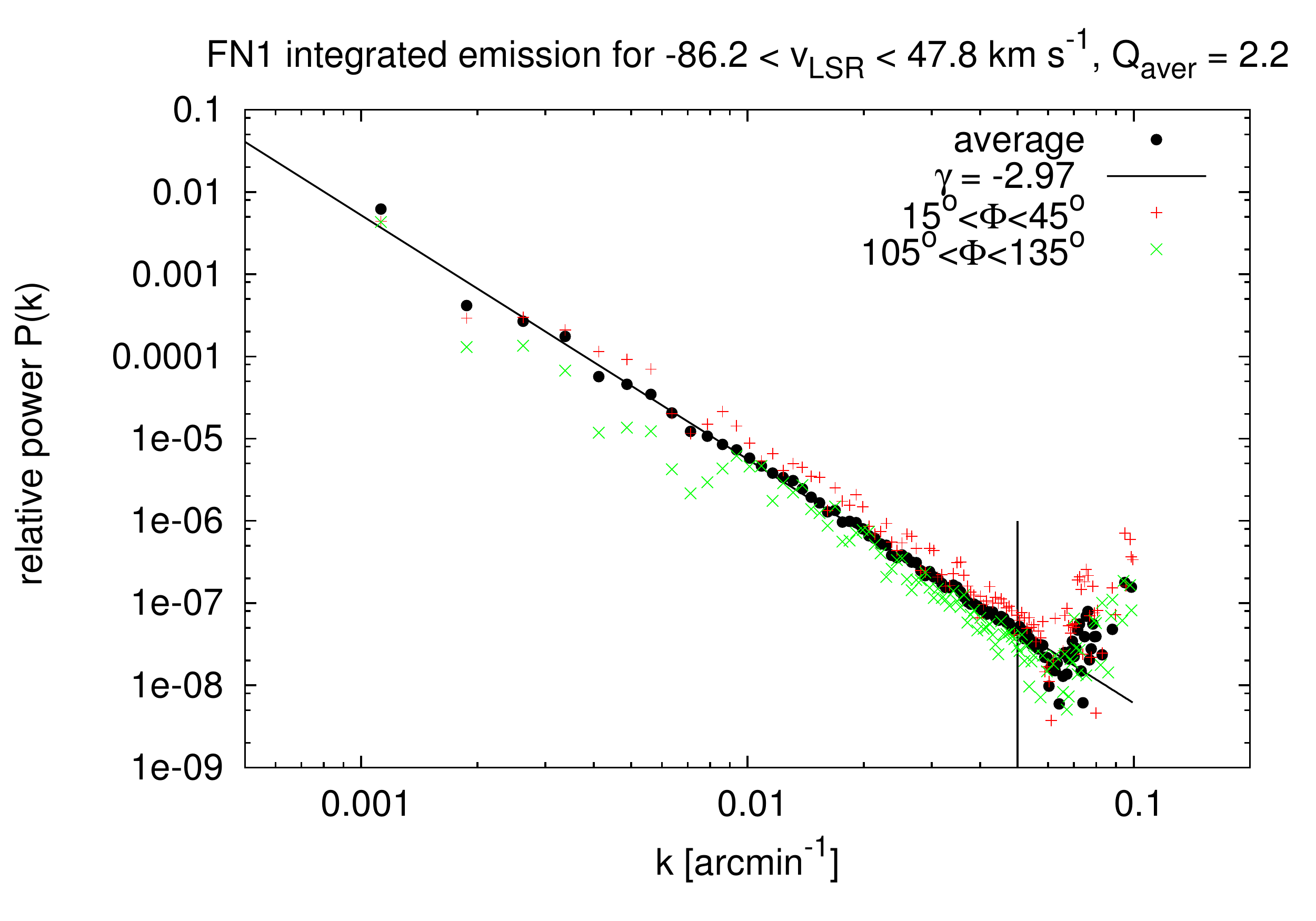}
   \caption{Average power spectrum for the integrated emission and
     fit power law (black) with $\gamma = -3.00 \pm 0.03 $ for $ k <
     0.05$ arcmin$^{-1}$ (vertical line). In addition the power
     spectrum for $30\degr < \Phi < 60\degr $ (red) and $120\degr < \Phi
     < 150\degr $ (green) is given. The average anisotropy factor is
     $Q_{\rm aver} = 2.0$. }
   \label{Fig_FN1_integrated}
\end{figure}

  Close to the peak of the average emission $T_{\rm B\, aver}$, but
  shifted by two channels to positive velocities we find a remakable
  narrow minimum in the spectral index distribution. One more channel
  shifted to positive velocities, at $v_{\rm LSR} = 3.97$ \kmss, there
  is a narrow peak for $Q_{\rm aver}$. We checked this channel in more
  detail and found anisotropies peaking at an angle of $ \phi \sim
  30\degr$.  The average power spectrum at this velocity is shown in
  Fig. \ref{Fig_FN1_74}, together with power spectra in direction of the
  derived position angle and perpendicular. There are marked
  anisotropies. In comparison to this single channel, representing a
  thin velocity slice, we find for the velocity centroid and the average
  emission, both determined for $-86.2 < v_{\rm LSR} < 47.8$ \kms, only
  insignificant anisotropies $Q_{\rm aver} = 2.2$
  (Figs. \ref{Fig_FN1_centroid} and \ref{Fig_FN1_integrated}).

\begin{figure}[tbp]
   \centering
   \includegraphics[width=6.5cm,angle=-90]{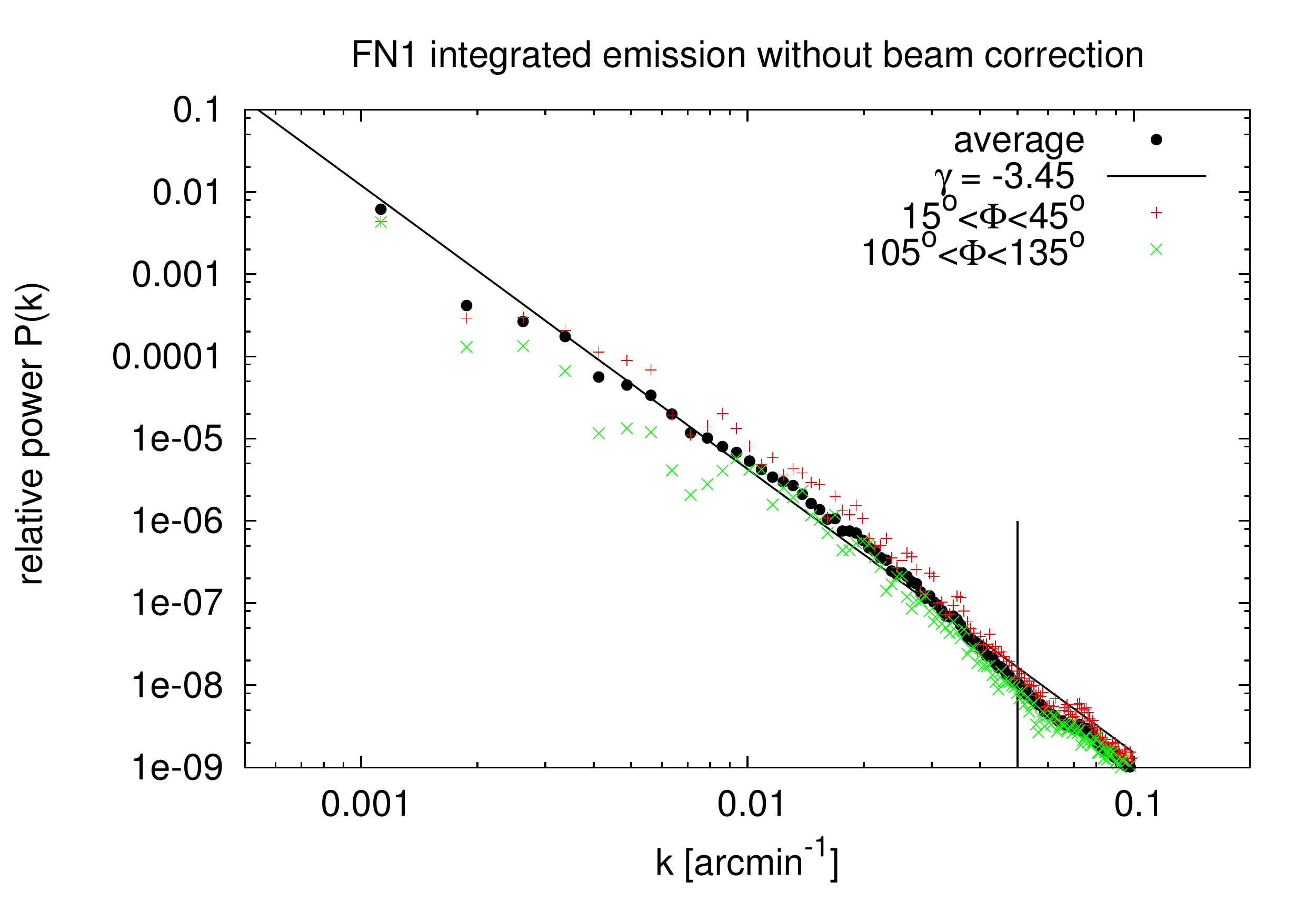}
   \caption{To demonstrate biases in power distribution and derived
     spectral indices we repeat the analysis from Fig
     \ref{Fig_FN1_integrated} without a beam correction. The power
     spectrum at high spatial frequencies is biased.}
   \label{Fig_XFN1_integrated}
\end{figure}

The spectral index for the velocity centroid is $\gamma = -2.80 \pm
0.03 $ and $\gamma = -2.97 \pm 0.02 $ for the average emission profile.
\citet{Miville-Deschenes2007} find for centroid and integrated
emission $\gamma = -3.4 $, inconsistent with our result. Their analysis
might suffer from systematic biases, causing a steep spectral
index. They deconvolved the LDS power distribution assuming a FWHM beam
width of 30\arcmin while the telescope beam is 36\arcmin.  When
generating maps there is some additional smoothing, depending on the
methods used. Last not least, the LDS is not Nyquist sampled; this
introduces some additional biases that cannot easily be specified. 

We repeated our spectral analysis for the integrated emission omitting
the beam correction. The result is plotted in
Fig. \ref{Fig_XFN1_integrated} and should be compared with
Fig. \ref{Fig_FN1_integrated}. The distribution at high spatial
frequencies with a spectral steepening is a fraud, though it appears to
be less noisy.  When omitting a beam correction, the slope steepens from
$\gamma = -2.97 $ to $\gamma = -3.45 $. We see that a subtle lapse in
data processing may lead to significant biases. Anisotropies at high
spatial frequencies are suppressed.  After reviewing the literature
cited in Sect. \ref{intro} we note that it is not obvious whether in all
cases a proper beam correction has been applied to the data. Some of the
steep spectral indices might be due to this systematic bias, also a
steepening of the power spectrum should be carefully investigated. 

Recently \citet{Blagrave2016} reported a similar problem in in the Ursa
Major region. \citet{Miville-Deschenes2003} derive there a power law
index $\gamma = -3.6 \pm 0.2$ for both, column density map as well as
velocity centroid, while \citet{Blagrave2016} measured $\gamma = -2.68
\pm 0.14$ for column densities and $-2.54 < \gamma < -2.42 $ for the
velocity centroid. We tested this case with EBHIS data and obtained for
column densities and velocity centroid $\gamma = -2.68 \pm 0.07$,
consistent with \citet{Blagrave2016}.

Our conclusion after calculating power spectra and anisotropies for the
FN1 field is that the 3C~196 anisotropies at $v_{\rm LSR} \sim -3.76$
\kmss are outstanding. For $ -11.5 < v_{\rm LSR} < -1.2 $ \kmss all
filaments are aligned with the magnetic field. Anisotropies for the
3C~196 field at $v_{\rm LSR} \sim 11$ \kmss as well as for the two
comparison fields are mostly around $Q_{\rm aver} \sim 2$. However even
in the FN1 field we find channels with strong anisotropies. The major
difference is that in this case the anisotropies are restricted to
narrow velocity intervals and the position angles change
systematically. Checking the dust emission observed by {\it Planck} at
353 GHz we find a pretty weak and noise limited signal without any
indications for filamentary structures. Thus there is no clear evidence
that one of the velocity channels is associated with a dust filament in
this region. The power distribution at $v_{\rm LSR} = 3.97$ \kmss will
be discussed later in more detail. As long as further studies about
anisotropies in the \hi are missing, we consider $Q_{\rm aver} \sim 2$
as typical for the unperturbed \hi distribution. As ``unperturbed'' we
understand regions that are not affected by obvious CNM filaments that
can be seen in USM maps.

\section{Discussion}
\label{Discussion}

After characterizing anisotropies for an unperturbed \hi distribution,
we return to a discussion of the 3C~196 field. We also compare
observations with predictions from theoretical models. 

To compare the derived power law index and the
anisotropy measure $Q$ with anisotropies as expected for MHD turbulence,
we first need to relate $Q$ to the wave vector anisotropy in terms of
turbulence theories. These are usually based on the analysis of the
structure of the eddies as a function of scale in presence of a local
magnetic field. Parallel and perpendicular to the magnetic field eddies
become correlated. As the turbulent cascade proceeds to larger spatial
frequencies the eddies become more and more elongated along the
direction of the magnetic field. The deformation increases until
dissipation is affecting the turbulent flow \citep{Goldreich1995} for
recent reviews see \citet{Cho2003,Brandenburg2013}.  

For a description how eddies (or wave packets) are related to the
corresponding wave vectors see \citet[][Fig. 6]{Cho2000}. For a wave
packet with extension $L_1$ perpendicular to the local magnetic field
direction and $L_2$ parallel to the field, the corresponding wave
vectors in Fourier space are for $L_1$ transformed to $1/k_{\perp}$ and
for $L_2$ respectively to $1/k_{\parallel}$. Anisotropy implies in the
Fourier domain isophotes with scales $k_{\parallel} < k_{\perp}$. The
power spectra $P_{\parallel}$ and $P_{\perp}$ are defined
correspondingly, $P_{\parallel}(k) < P_{\perp}(k)$. 

According to \citet{Goldreich1995}, turbulence spectra $P_{\perp}$ are
in Alfv{\'e}n and slow modes nearly unaffected by the magnetic field
with power spectra close to the isotropic case.

\subsection{Field direction and spectral index }
\label{Power}

\begin{figure}[tbp]
   \centering
   \includegraphics[width=6.5cm,angle=-90]{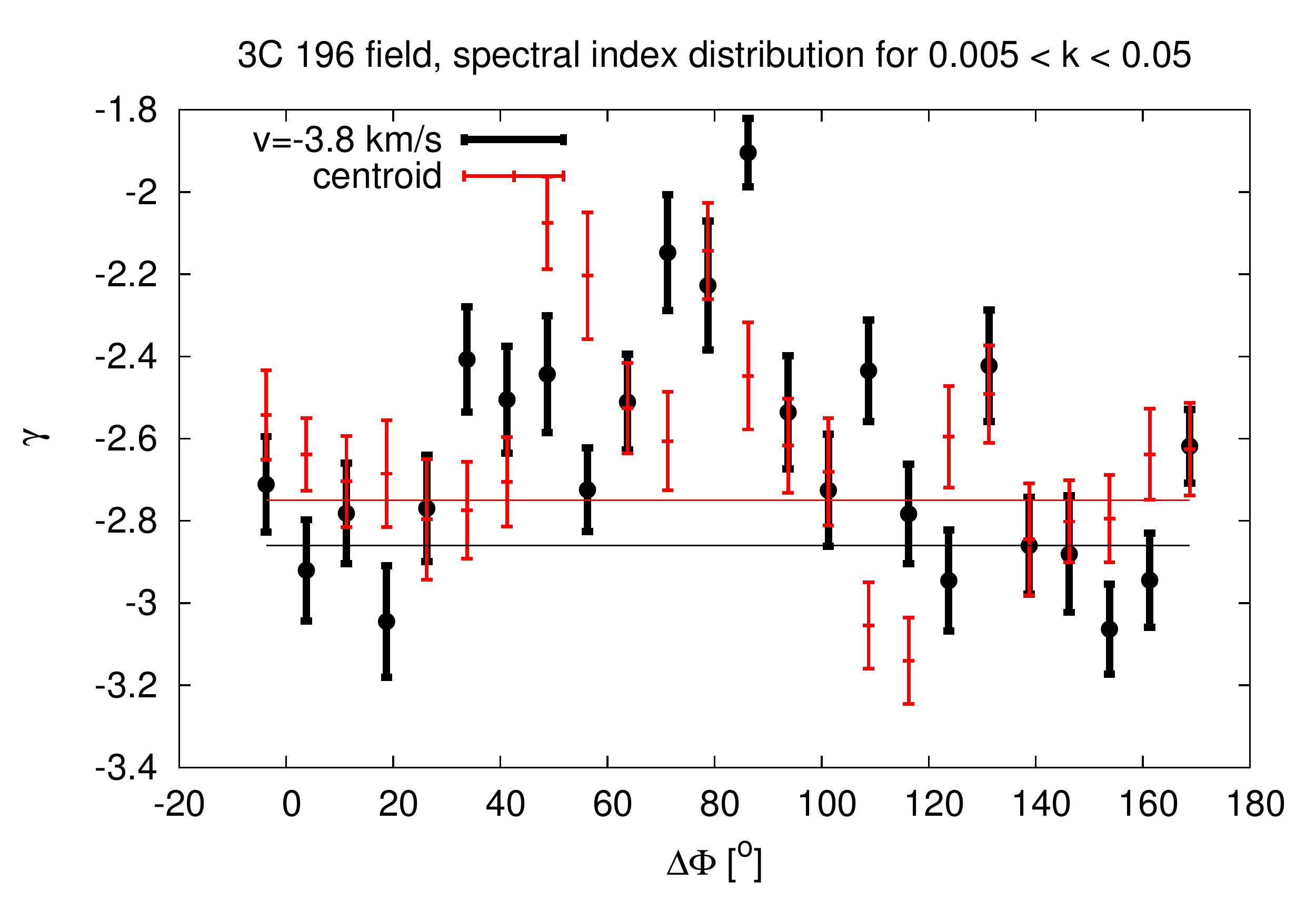}
   \caption{Distribution of spectral index $\gamma$ with respect to the
     field angle $\Delta\Phi = \Phi - \Phi_{\perp}$.  Fits for
     individual sectors with width $2\delta\Phi = 7\fdg5$ (black) are
     shown for the channels at $ v_{\rm LSR} = -3.76 $ \kmss and the
     velocity centroid (red). Only spatial frequencies $0.005 < k <
     0.05$ arcmin$^{-1}$ are considered. Isotropic spectral indices are
     plotted with horizontal lines for comparison. }
   \label{Fig_plot_fit}
\end{figure}

We determined spectral indices $\gamma$ with respect to the field angle
$\Delta\Phi = \Phi - \Phi_{\perp}$, restricting the fit to spatial
frequencies with clear anisotropies, $0.005 < k < 0.05$. Figure
\ref{Fig_plot_fit} shows the results for two cases, the channel at $
v_{\rm LSR} = -3.76 $ \kmss (black) and the velocity centroid (red). We
find in both cases a complex distribution but a general trend for a
steep power index in field direction, consistent with the mean
(isotropical) spectral indices indicated with the horizontal lines in
Fig.  \ref{Fig_plot_fit}. Parallel to the magnetic field, for $\Delta
\Phi \sim 90\degr$, the spectral indices flatten, approaching $\gamma
\sim -2.1 $.  \citet{Goldreich1995} predict for $P_{\perp}$ that the
spectral index for anisotropic turbulence is nearly identical to the
isotropic case as indicated by the horizontal lines in
Fig. \ref{Fig_plot_fit}. This prediction is in good agreement with our
result.

\subsection{Anisotropies $Q(v_{\rm LSR},k)$}
\label{Anisotropies}

From observations, $Q_{\rm aver}$ appears far more useful to
characterize anisotropic turbulence than fitting spectral indices to
power spectra. Here we try to go a step further to find out whether
$Q(k)$ can be used to determine systematic changes in spectral indices.

Figure \ref{Fig_3C196_angle} shows that anisotropies have a constant
position angle for velocities $-11.5 < v_{\rm LSR} < -1.2$ \kms, also
$Q_{\rm aver}$ is well defined in this range
(Fig. \ref{Fig_overview}). We choose spectral channels that may be
considered as independent and typical for this range. In
Fig. \ref{Fig_chan_aniso_35} we display $Q(k)$ for the channel map at $
v_{\rm LSR} = -6.3 $ \kmss (black dots), $ v_{\rm LSR} = -8.9 $ \kmss
(blue), and for the velocity centroid (red). There is quite some scatter
but we obtain a general increase of the anisotropies up to $ k \sim
0.025 $ arcmin$^{-1}$. For $ 0.025 \la k \la 0.07$ arcmin$^{-1}$
(enclosed by horizontal lines) the anisotropies appear to decay, the
highest spatial frequencies are noise dominated and cannot be
interpreted. For the rising branch we fit for each of our samples a
power law $Q(k) \propto k^{\beta}$ and obtain $ 0.59 \la \beta \la 0.81
$. We find clear evidence that anisotropies increase with spatial
frequencies as predicted by \citet{Goldreich1995} but the slope appears
not to be very well defined. The observed data are affected by local
effects that cause quite some scatter.

\subsection{Decay of anisotropies? }
\label{Decay}

\begin{figure}[tbp]
   \centering
   \includegraphics[width=6.5cm,angle=-90]{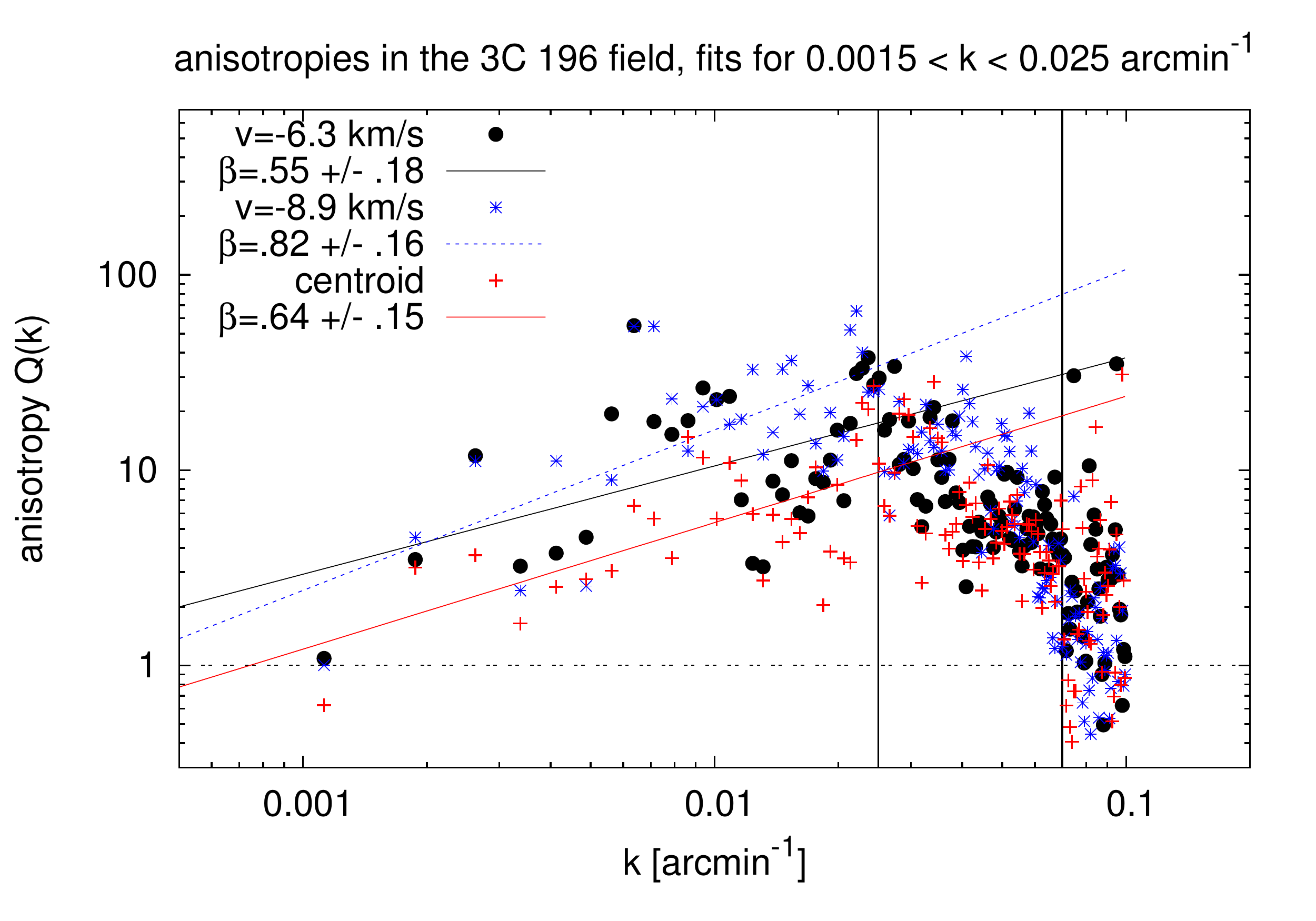}
   \caption{Anisotropies $Q(k)$ for the channels at $ v_{\rm LSR} = -6.3
     $, -8.9 \kmss (black and blue), and for the velocity centroid
     (red).  Power laws $Q \propto k^{\beta}$ were fit for $k < 0.025$
     arcmin$^{-1}$.}
   \label{Fig_chan_aniso_35}
\end{figure}

For $ k > 0.025$ arcmin$^{-1}$ we find in Fig. \ref{Fig_chan_aniso_35} a
general systematical decrease of $Q(k)$. This tail suggests a decay of
the observed anisotropies in the same sense as the energy decays in the
turbulent flow to progressively smaller scales. The question arises
whether this apparent decay may be caused by instrumental limitations.
A decay may be mimicked by noise limitations. We calculate the
signal-to-noise ratio (SNR) of the anisotropies by comparing
$P_{\parallel}(k)$ with the power in the matched noise template $N(k)$
(compare Fig. \ref{Fig_spec_37} with Fig. \ref{Fig_spec_aver}). For $ k
< 0.054$ arcmin$^{-1}$ we find a SNR of three or better. The SNR for
$P_{\perp}(k = 0.054)$ is more than five times better. Hence $Q(k)$ is
for $ k < 0.054$ arcmin$^{-1}$ significant and unaffected by
instrumental problems.  Numerical uncertainties may also cause
limitations. We use the Singleton FFT algorithm in double precision and
are not aware about accuracy limitations.

\begin{figure}[tbp]
   \centering
   \includegraphics[width=6.5cm,angle=-90]{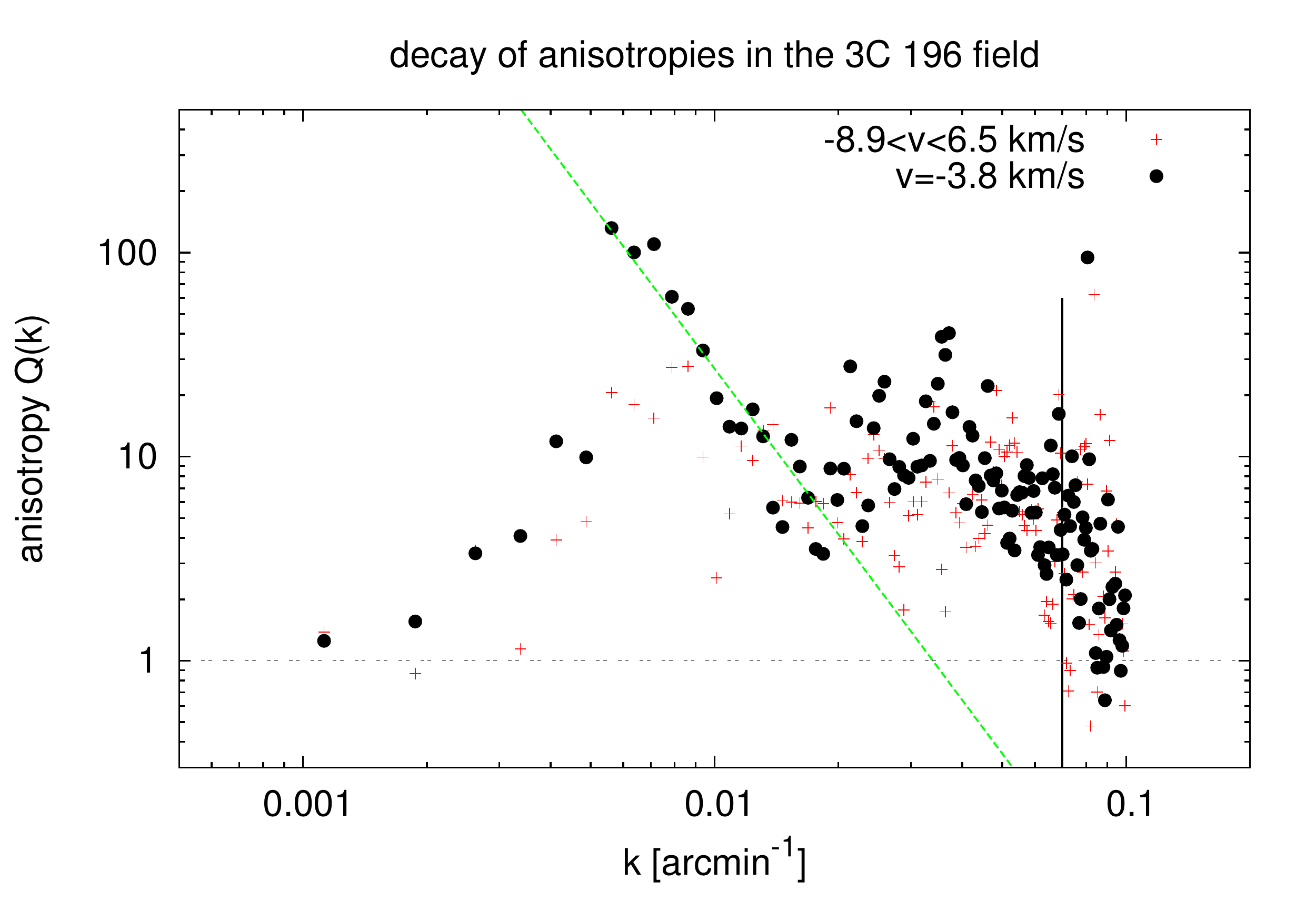}
   \caption{Anisotropies $Q(k)$ for the channel at $ v_{\rm LSR} = -3.76
     $ \kms (black dots). The strong local anisotropy of $Q \sim 130$ at
     $ k = 0.0056 $ arcmin$^{-1}$ decays as $Q \propto k^{-2.7}$. Also
     the anisotropies at $k > 0.04$ arcmin$^{-1}$ appear to decay in a
     similar way. For comparison we show thick slice anisotropies for
     $-8.9 < v_{\rm LSR} < 6.5$ \kmss with red crosses.  }
   \label{Fig_chan_37_aniso}
\end{figure}

\begin{figure}[tbp]
   \centering
   \includegraphics[width=6.5cm,angle=-90]{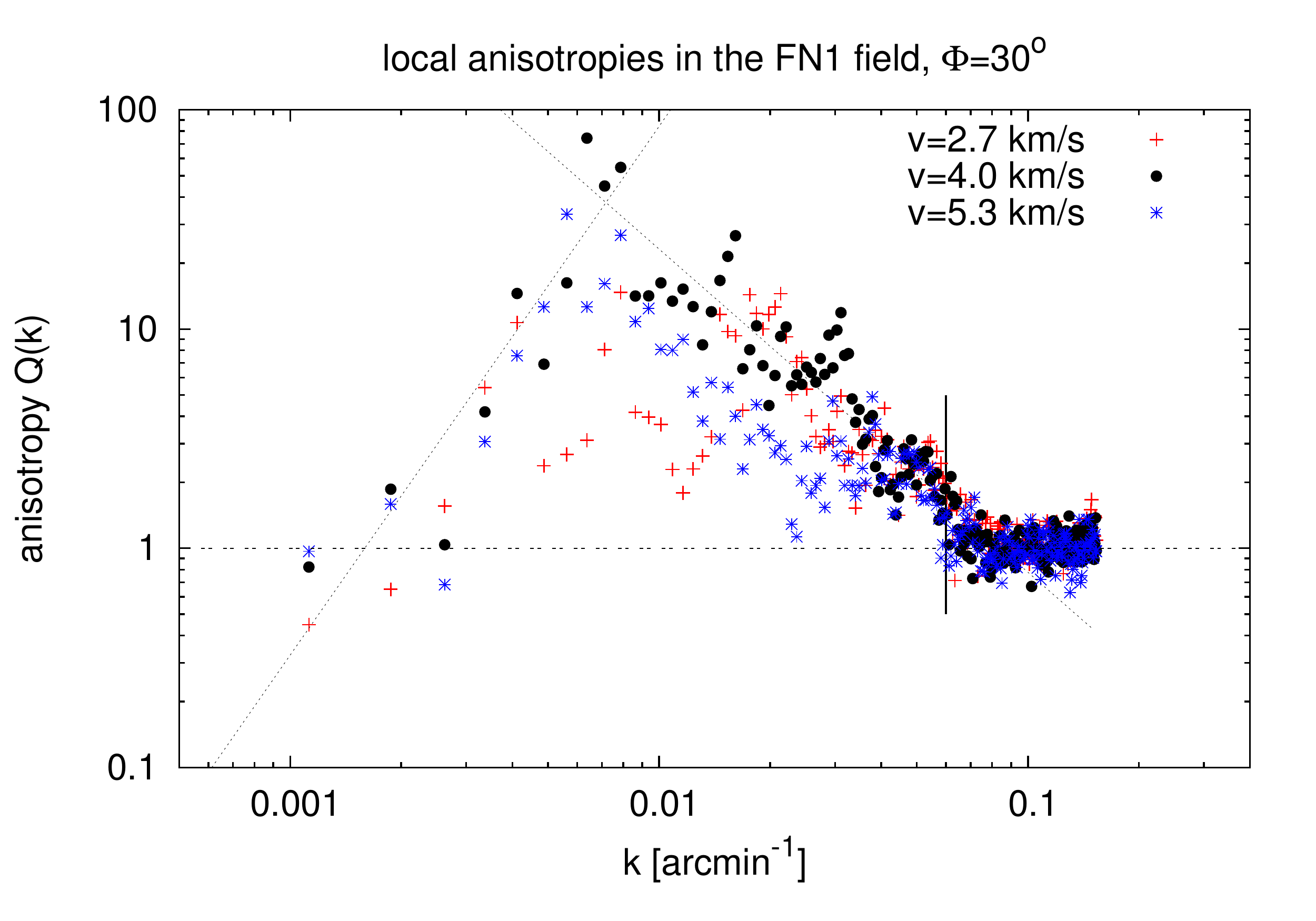}
   \caption{Anisotropies $Q(k)$ in the FN1 field for the channel at $
     v_{\rm LSR} = 4.0 $ \kmss (black dots) and neighbor channels at $
     v_{\rm LSR} = 2.7 $ \kmss (red) and $ v_{\rm LSR} = 5.3 $ \kmss
     (blue). Anisotropies at $ v_{\rm LSR} = 4.0 $ \kmss appear to rise
     with a spectral index of $\gamma \sim 2.4 $ and decay with $\gamma
     \sim -1.6 $ (dotted lines).  }
   \label{Fig_spec_Q_FN1_74}
\end{figure}

Common in Fig. \ref{Fig_chan_aniso_35} is that the increase of the
anisotropies $Q(k)$ is accompanied with quite some scatter. This scatter
is much stronger then the expected random noise and appears to indicate
local but systematical deviations caused by external forces. The channel
map at $ v_{\rm LSR} = -3.76 $ \kms, shown in Fig.
\ref{Fig_chan_37_aniso}, has the strongest local scatter with the
strongest anisotropy of $Q \sim 130$ at $ k = 0.0056 $
arcmin$^{-1}$. Considering the turbulent flow from large eddies to
smaller ones, this outstanding anisotropy pops suddenly up (by a factor
of ten), is highly significant but inconsistent with the typical
anisotropies that are observed at other velocities. Towards high spatial
frequencies this anisotropy decays as $Q \propto k^{-2.70 \pm .24}$
(green line in Fig. \ref{Fig_chan_37_aniso}), a slope that is in good
agreement with the 2D Kolmogorov index.

It may be fortuitous to relate the decaying part of the anisotropies to
the Kolmogorov index. Figure \ref{Fig_spec_Q_FN1_74} shows that the
strongest anisotropy in the FN1 field has a comparable pattern with a
rising part at low spatial frequencies up to $Q = 74$ at $k = 0.0064
$. This branch can be fit with a power law index $\gamma = 2.4 \pm
0.4$. The anisotropies appear to decay for larger spatial frequencies
with a power law index $\gamma = -1.6 \pm 0.1$. These spectral indices
differ significantly from the results obtained for 3C~196, possibly
indicating different processes causing the anisotropies. The 3C~196
filaments are clearly aligned with the magnetic field but using 353 GHz
{\it Planck} data for the FN1 field it is hardly possible to relate the
filament to a magnetic field direction. The average magnetic field
position angle for an area with a diamenter of 10\degr is $\phi_{\rm
  mag} = 111\fdg4 \pm 2\fdg2$ and $\phi_{\rm mag} = 106\fdg4 \pm 1\fdg0$
respectively for a diamenter of 20\degr (F. Boulanger, private
communication). Different processes may cause different spectral
indices.

The strong FN1 anisotropy is caused by CNM with a geometric mean Doppler
temperature of $T_{\rm D} \sim 250 $ K. Due to the low temperature,
neighbor channels are nearly uncorrelated and show significantly
different anisotropies. The position angles show significant changes
(Fig. \ref{Fig_FN1_angle}), it is not possible to assign an unambiguous
relation between filament and magnetic field direction. We found peak
anisotropies $ \phi \sim 30\degr$ for $ v_{\rm LSR} = 3.97 $ \kms. The
velocity channel at $ v_{\rm LSR} = 10.4 $ \kmss has $\phi = 9.3\degr
\pm 13\degr $ with a peak anisotropy of $Q = 55$, the channel at $
v_{\rm LSR} = -1.2 $ \kms has $\phi = 40\degr \pm 11\degr $ with a peak
anisotropy of $Q = 10$ respectively. In both cases we derive Doppler
temperatures $T_{\rm D} \la 350 $ K, the \hi gas belongs to the CNM.
\citet{Miville-Deschenes2007} argue that 2/3 of the column density in
the FN1 region is made of WNM and 1/3 of thermally unstable gas ($T \sim
2600$ K). They further find that this field contains a signature of CNM
type gas at a very low level which could have been formed by a
convergent flow of WNM gas. Details of the CNM component are not
specified but we conclude that this gas probably corresponds to the
component that shows anisotropies.

The 3C~196 filament showing strong anisotropies at $ v_{\rm LSR} =
-3.76 $ \kmss is aligned with the magnetic field and is of
particular interest since the CNM \hi distribution at this velocity
shows the best visible correlation of all \hi USM channel maps with the
LOFAR filaments, see Fig. \ref{Fig_cubep_USM}. The strong local
anisotropy is also visible in Fig. \ref{Fig_spec_37}. Perpendicular to
the magnetic field (red) the spectral power increases locally with $k$
while there is a stagnation in parallel direction (green). We interpret
this as an indication for a strong local interaction.

Assuming that the observed anisotropies are caused by MHD turbulence, we
might consider the case whether the magnetic field could cease at $ k
\ga 0.025$ arcmin$^{-1} $. It is difficult to prevent the magnetic field
from small scales but from the deformation of eddy shapes, reconnection
appears to be a possibility.  \citet[][Eq. A13]{Lazarian1999} estimated
for this case under typical CNM conditions a scale of 0.03 pc. For
dissipation effects, the only known mechanism in the ISM that leads to
comparable scales is the ion-neutral friction. \citet{Hennebelle2013}
estimate the length associated to the ion-neutral friction in the
typical CNM to be in the range $ 3~10^{-3}$ to $ 2~10^{-2}$ pc. We may
compare these scales with estimates of the filament thickness from
observations. For unresolved filaments, assumed to be sheets seen
edge-on, \citet{Kalberla2016} estimate a median thickness of 0.3 pc, and
in case that magnetic pressure confinement applies, a thickness of 0.1
pc.  \citet{Naomi2006} and \citet{Clark2014} obtained comparable
results. 

For local \hi filaments at a distance of 100 pc, based on the
median distance to the wall of the local cavity \citep[][from color
excess measurements]{Lallement2014}, the break in the anisotropy
distribution at $ k \sim 0.025$ arcmin$^{-1}$ corresponds to an eddy
extension of $l \sim 1.2$ pc perpendicular to the magnetic field. This
is by far too large to explain a decay of anisotropies by reconnection
or dissipation by ion-neutral friction. A different explanation needs to
be found.

\subsection{Spectral index dependencies}
\label{Index}

We find in case of well defined anisotropies in the range $ 0.005 < k <
0.05 $ arcmin$^{-1}$ always $Q(k) = P_{\perp}(k) /P_{\parallel}(k) \gg 1
$. The spectral index $\gamma_{\perp}$ for $P_{\perp}$ is well defined
and close to the isotropic index $\gamma_{\rm iso} \sim -8/3$, one of
the basic predictions by \citet{Goldreich1995}. For $P_{\parallel}(k)$
we obtain an average power index $ \gamma_{\parallel} = -2.06 \pm
0.2$ (Fig. \ref{Fig_plot_fit}). 

According to $Q(k) = P_{\perp}(k) /P_{\parallel}(k) \propto
k^{\gamma_{\perp}}/k^{\gamma_{\parallel}} = k^\beta $ this implies that
anisotropies increase with a spectral index $\beta \sim 2/3$, we fit $
0.59 < \beta < 0.81 $ (see Fig. \ref{Fig_chan_aniso_35}), consistent
with the second prediction by \citet{Goldreich1995}. 

However, anisotropies do not increase in an unlimited way. Our data
suggest that anistropies decrease for $ k \ga 0.025$ arcmin$^{-1}$.
Also the power distribution for the channel at $ v_{\rm LSR} = -3.76 $
\kmss does not agree with the predictions. At $ k = 0.0056 $
arcmin$^{-1}$ we find a strong anisotropy $Q(k) \sim 130$ which is
inconsistent with MHD model assumptions. This anisotropy decays
according to $Q(k) = P_{\perp}(k) /P_{\parallel}(k) \propto
k^{\gamma_{\perp}}/k^{\gamma_{\parallel}} = k^\beta \sim k^{-2.7}$. For
this channel we measure $\gamma_{\rm iso} \sim \gamma_{\perp} \sim
-2.86$, implying $\gamma_{\parallel} \sim 0.16 $. Within the errors the
power distribution parallel to the field is flat, $P_{\parallel} =
const$. The flat part of the power distribution can be seen in
Fig. \ref{Fig_spec_37} (green) for $ 0.0056 < k < 0.02 $ arcmin$^{-1}$.
Quite surprisingly, we observe for the affected range in spatial
frequencies a signal that is completely uncorrelated. For
$P_{\parallel}(k)$ the turbulent cascade is suspended, there is no
turbulent energy transport from larger to smaller eddies. This
interpretation assumes that the $ P_{\perp}$ distribution is largely
unaffected by the decay and close to $ P_{\rm iso}$ as predicted by
\citet{Goldreich1995}.

\subsection{Eddy shapes}
\label{Eddy}

\begin{figure}[tbp]
   \centering
   \includegraphics[width=6.5cm,angle=-90]{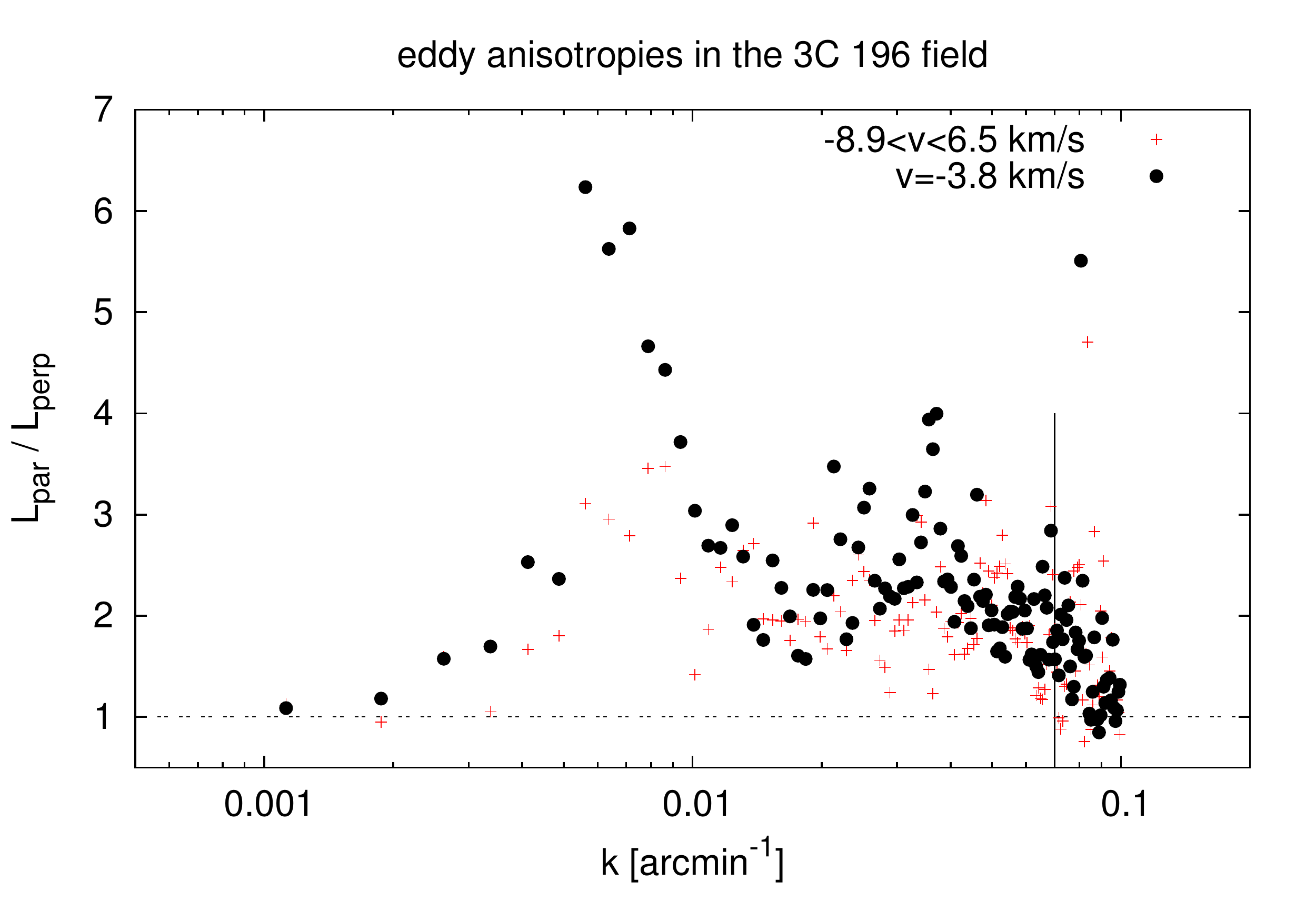}
   \caption{Eddy anisotropies $L_{\parallel}/L_{\perp} \propto
     Q(k)^{3/8}$ for the channel at $ v_{\rm LSR} = -3.76 $ \kms (black
     dots). Thick slice eddy anisotropies for $-8.9 < v_{\rm LSR} < 6.5$
     \kmss are added for comparison (red crosses).  }
   \label{X_Fig_chan_37_aniso}
\end{figure}

For local \hi filaments we estimate a distance of 100 pc, based on
\citet{Lallement2014}. The filaments extend approximately over 20\degr
(see Fig. \ref{Fig_HI_gal}), corresponding to an outer turbulence scale
of $l_o \sim 35$ pc. The break in the anisotropy distribution at $ k
\sim 0.025$ arcmin$^{-1}$ corresponds for this distance to an average
eddy extension of 40\arcmin or $l \sim 1.2$ pc perpendicular to the
magnetic field. For a power law index of $\gamma = -8/3$ the wave vector
anisotropy is $k_{\perp} / k_{\parallel} \sim Q^{3/8} \sim 2.4$
\citep{Chen2010}. Hence the dominant eddy size is 1.2 pc by 2.8 pc. For
$Q = 130$ we get accordingly an eddy shape that is typically 6.2 times
more extended along the field line than perpendicular. The size is here
5.2 pc by 32 pc.  In Fig. \ref{X_Fig_chan_37_aniso} we display eddy
anisotropies $L_{\parallel} / L_{\perp} $ as function of spatial
frequency.

\begin{figure}[tbp]
   \centering
   \includegraphics[width=6.5cm,angle=-90]{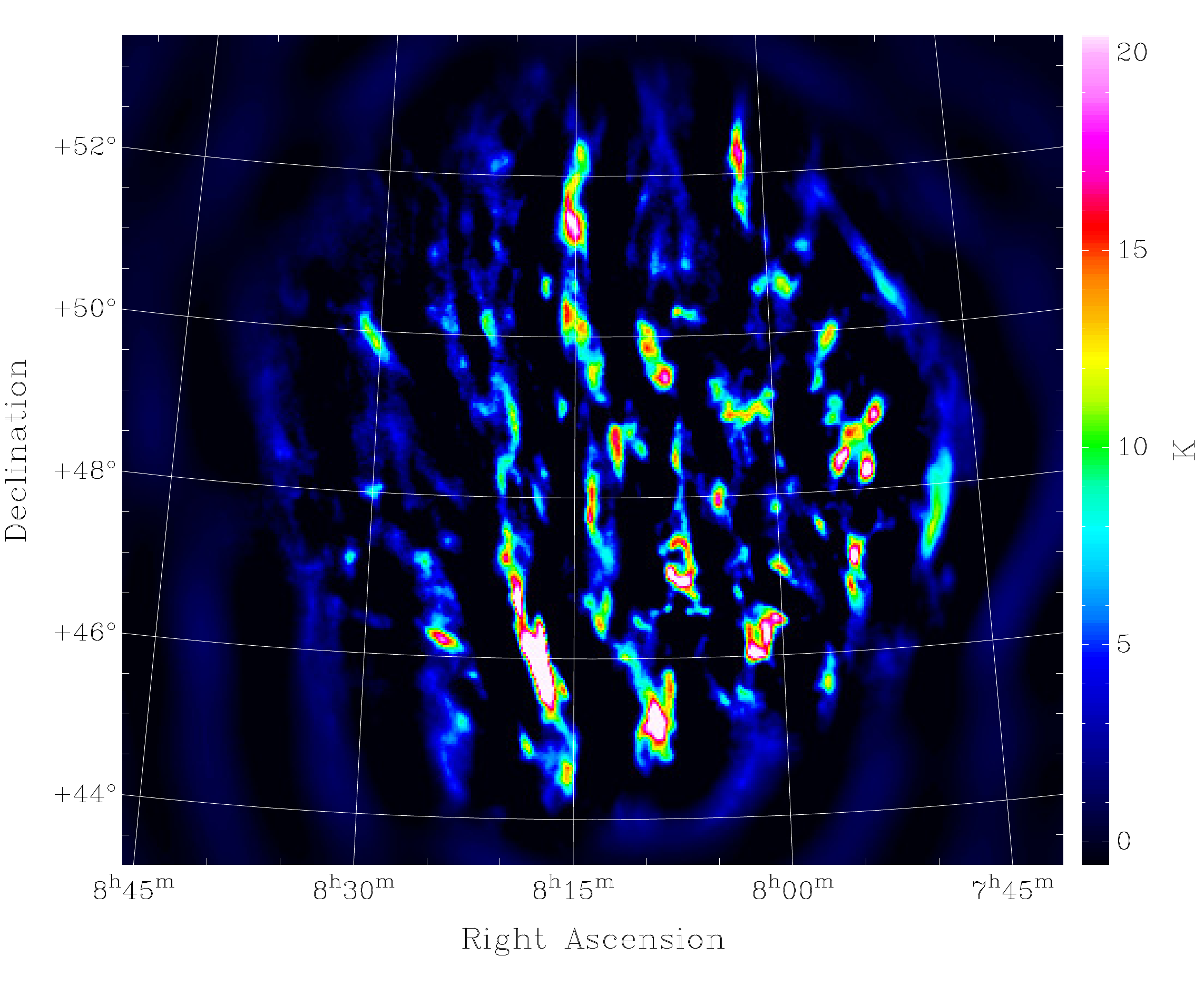}
   \caption{Apodized brightness temperature distribution in direction to
     3C~196 at $v_{\rm LSR} = -3.76$ \kmss for spatial frequencies $k >
     0.0056$ arcmin$^{-1}$ with anisotropies $ Q < 130$. }
   \label{Fig_Im37_mask}
\end{figure}

\begin{figure}[tbp]
   \centering
   \includegraphics[width=6.5cm,angle=-90]{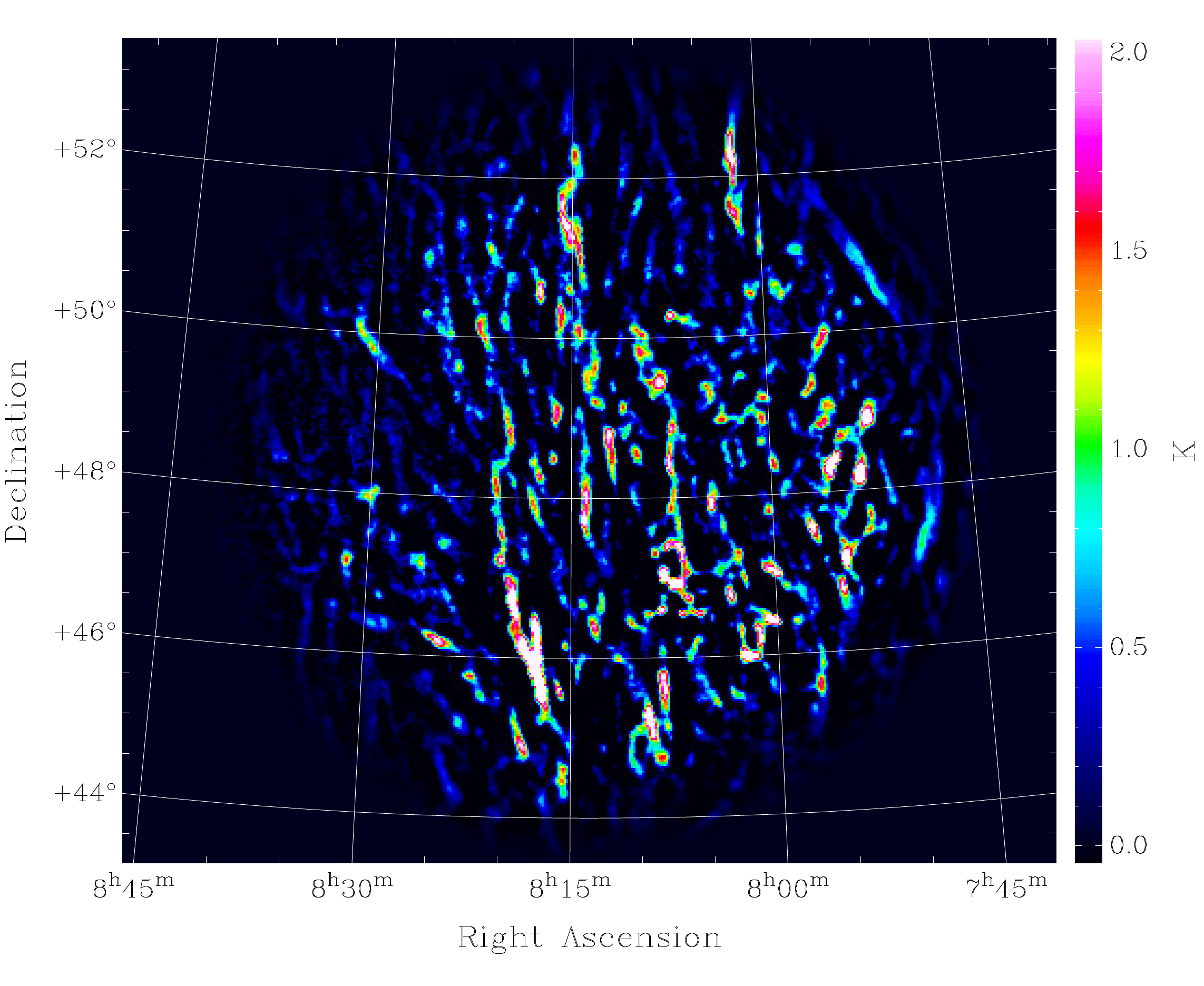}
   \caption{Apodized USM temperature map for the CNM in direction
     to 3C~196 at $v_{\rm LSR} = -3.76$ \kms, demonstrating that the CNM
     is dominated by filaments. Anisotropies at high spatial frequencies
     can be inferred from CNM maps.  }
   \label{Fig_Im37_USM}
\end{figure}

In Fig. \ref{Fig_Im37_mask} we display data for the channel at $v_{\rm
  LSR} = -3.76$ \kmss for spatial frequencies $k > 0.0056 $
arcmin$^{-1}$. From our data this is the most extreme anisotropy and
probably the best possible visualization of what one may consider as
eddies or wave packets. For comparison we plot in
Fig. \ref{Fig_Im37_USM} the CNM gas, indicating that these eddies
coincide with rather cold \hi gas. In this case we derive the average
anisotropies $Q_{\rm aver} = 11$ but the peak anisotropy reaches $Q =
65$, only 50\% compared to the \hi channel map in
Fig. \ref{Fig_Im37_mask}. When generating USM maps, spatial frequencies
$ k \la 0.033$ arcmin$^{-1}$ are attenuated including the anisotropies
observed at $ 0.01 < k < 0.03 $ arcmin$^{-1}$. Due to the fact that MHD
driven anisotropies increase with spatial frequency, USM maps give a
good first order approximation of the anisotropic distribution of the
\hi at high spatial frequencies.  The visual inspection of the USM map
implies that most of the filaments are unresolved.

\section{Summary and conclusions}
\label{Summary}

It was recently shown that a major part of the CNM is organized in cold
filaments with median Doppler temperatures $T_{\rm D} \sim 223$ K that
are aligned with the magnetic field \citep{Kalberla2016}. These
structures can be highly anisotropic with aspect ratios up to $\sim 400$
\citep{Heiles2005,Naomi2006,Clark2014,Kalberla2016}. Such a medium is
expected to be driven by MHD turbulence and asymmetries should have a
large impact on the dynamics of the turbulent cascade.

To determine anisotropies in MHD driven turbulence we decided to analyze
the \hi distribution in direction to 3C~196. This field is one of
the primary fields of the LOFAR-Epoch of Reionization key science
project. Linear polarization structures have been observed with LOFAR by
\citet{Jelic2015}.  Strikingly straight filaments
were found to be aligned with the magnetic field \citep{Zaroubi2015} and
\hi shows a similar filamentary structure (Fig. \ref{Fig_cubep_USM}). 

We find that average power spectra, as well as power spectra 
perpendicular to the magnetic field, are for thin slices in velocity
close to the expected Kolmogorov index $\gamma = -8/3$.  The general
impression is that in case of anisotropies the parallel and
perpendicular power spectra appear to the first order in logarithmic
presentation just ``offset'' from the isotropic distribution, see
Figs. \ref{Fig_spec_angle} to \ref{Fig_spec_49}, \ref{Fig_spec_33_45},
and \ref{Fig_spec_33_45_centroid}. More important are however local and
correlated deviations from the isotropic power distribution. The data
show quite some fluctuations which we interpret as caused by external
forces.

The \hi distribution in direction to 3C~196 has two major components
(Fig. \ref{Fig_overview}), at $v_{\rm LSR} \sim -1 $ \kmss and $ \sim
13$ \kmss but only the component at $v_{\rm LSR} \sim -1 $ \kmss appears
visually to be related to the LOFAR filaments.  We observe strong
average anisotropies, $ 10 < Q_{\rm aver} < 20 $, in the spectral power
distribution of the $v_{\rm LSR} \sim -1 $ \kmss component. The $v_{\rm
  LSR} \sim 13$ \kmss component does not show significant
anisotropies. Thus anisotropies are clearly linked to \hi filaments that
appear visually to be correlated with LOFAR filaments.  The analysis of
two more comparison fields show that anisotropies, unrelated to
filamentary structures, are mostly restricted to $Q_{\rm aver} \sim 2$.

The derived anisotropy position angle (Fig. \ref{Fig_spec_angle}) agrees
well with the orientation of the LOFAR filaments \citep{Jelic2015} and
the direction of the associated magnetic field, derived from {\it
  Planck} 353 GHz data \citep{Zaroubi2015}. The anisotropies are caused
by cold CNM filaments, we derive a geometric mean Doppler temperature
$T_{\rm D} \sim 161 $ K. The power law index $\gamma$ tends to be
steepest close to the minima of the Doppler temperature. Anisotropies $Q_{\rm
  aver}$ are not correlated with Doppler temperature or average line
emission, also we found no correlation with the power law index
$\gamma$.  Anisotropies lead to significant changes in the power
distribution, $P_{\parallel}(k) < P_{\perp}(k)$. Typical ratios reach
$Q(k) = P_{\perp}(k)/P_{\parallel}(k) \sim 10 $ to 20, peaking at $Q(k)
\sim 130$. Anisotropies increase with spatial frequency, $Q(k) \propto
k^{\beta}$ with $ 0.59 \la \beta \la 0.81 $ until $k \sim 0.025$
arcmin$^{-1}$.

We observe a decay of anisotropies at high spatial frequencies $ k \ga
0.025$ arcmin$^{-1}$.  From theory, MHD anisotropies should increase with
spatial frequency until a scale is reached where the turbulent flow
becomes dissipative. However we find no indications for a turnover in
the isotropic spectral distribution that might be considered as an
indication for dissipation. The average power distribution remains
unchanged. All we notice is a systematic decrease of the anisotropies.
Also the angular power spectrum of integrated polarized emission,
derived by \citet{Jelic2015} from LOFAR data, shows no change in
spectral slope down to the resolution limit at a scale of 4 armin.

A similar anisotropy decay with a spectral index of $ \gamma \sim -2.7$
is observed for the extreme case $Q \sim 130$ in the 3C~196
  field. While the power law index $\gamma_{\perp}(k)$ is almost
unaffected by the decay we find here within the errors
$\gamma_{\parallel}(k) \sim 0$, implying for the decay branch that there
is no turbulent energy transport for $P_{\parallel}(k)$. The turbulent
cascade appears to be suspended for $P_{\parallel}(k)$ until $k \sim
0.02$ arcmin$^{-1}$ with an anisotropy of $4 \la Q \la 8$, typical for
the other velocity channels, is reached. In other words, this process is
terminated when anisotropy according to the relation $Q(k) \propto
k^{2/3}$ \citep{Goldreich1995} is reached.

Our interpretation for the $ Q\sim 130$ case is that for the scale
length (eddy size) under consideration the enhanced power $P_{\perp}(k)$
was forced by an event that affected the eddy shape in a sense that
power from $P_{\parallel}(k)$ was transferred to $P_{\perp}(k)$
(compression of the eddy). After the external force is removed,
the turbulent cascade continues for a while in an isotropic way; the
eddy shapes need to relax according to the $Q(k) \propto k^{2/3}$
condition.

Our interpretation of the 3C~196 anisotropies may be fortuitous, the
processes causing these anisotropies are not yet understood. In case of
our comparison field, the FN1 region, we observe for a narrow velocity
interval another strong anisotropy (peak at $Q = 74$) but with
significant different spectral indices concerning the rise and decay of
anisotropies with spatial frequency (Fig. \ref{Fig_spec_Q_FN1_74}).

In our analysis we compare power spectra derived from thin and thick
slices as well as from centroid maps. The predicted steepening of power
indices \citep{Lazarian2000}, comparing thin and thick slices, is only
partly verified by us. We find that anisotropies are always best defined
for thin slices. Figure \ref{Fig_chan_37_aniso} demonstrates that thick
slices show less anisotropies and a further comparison with Fig. \ref
{Fig_chan_aniso_35} leads to the result that velocity centroids,
calculated for the same velocity range, may even be less suitable to
characterize anisotropies. Interestingly, our result contradicts one of
the main results by \citet{Kandel2016}, who finds that turbulence
anisotropies increase with velocity slice thickness.

The interpretation of our result is straightforward. Since the
filamentary CNM in the 3C~196 field is very cold ($T_{\rm D} \sim 160 $
K), adjacent velocity channels are uncorrelated for $\delta v_{\rm LSR}
\ga 3 $ \kms. On the other hand filaments in neighboring channels show
very similar structures. Filaments are probably \hi sheets seen
edge-on. Different slices, systematically offset from each other, are
observed at different velocities. These structures are not random but
correlated.  Integrating the data to obtain thick slices or velocity
centroids means for the analysis that local (uncorrelated) anisotropies
are smoothed out. Filaments, running parallel \citep[see Figs. 10, and
25 to 27 of][]{Kalberla2016}, are broadened artificially by averaging
and anisotropy is lost. The fluctuations caused by the CNM within a
single thin velocity slice is dominated by density effects but changes
in the observed density distribution with velocity are correlated
because of global anisotropies, probably influenced by the magnetic
field, that extend over a few channels.

For the FN1 field we find that position angles change systematically
with velocity. The CNM that is associated with the anisotropies is also
in this case sufficiently cold that neighbor channels are nearly
independent. Calculating the mean emission or the velocity centroid over
a range of velocities (in a thick slice), anisotropies smooth out and
loose significance. The result from our investigations is clearly that
anisotropies are associated with the CNM and not with the more extended
but diffuse WNM.

Alfv{\'e}n modes are incompressible and to the first order approximation
they do not create any density fluctuations. The expected power law
index is $\gamma = -11/3$ \citep{Kandel2016}. In Sect. \ref{Angular} we
excluded fast modes since these would cause eddies elongated
perpendicular to the magnetic field. The power spectrum in this case
would be isotropic with an index of $\gamma = -9/2$. For both modes this
is not what we observe. We conclude that we probably observe slow modes
in a particular favorable condition, the magnetic field is mostly
oriented perpendicular to the line of sight. According to
\citep{Heiles2005} the expected average field strength is $ 6~ \mu$G,
the turbulent $\beta \sim 0.3$, the Alfv{\'e}nic Mach number $M_{\rm A}
\sim 1.1$, and the Alfv{\'e}n velocity $v_{\rm A} \sim 1.5$ \kms.  LOFAR
observations and polarized dust emission by {\it Planck} restrict the
total magnetic field strength for the central filament to $ B_{\rm tot}
\la 6.5 \mu$G and the component along the line of sight to $ B \ga 1.2\
\mu$G. The pulsar J081558+461155, is located about $2\deg$ south of
3C~196. \citet{Jelic2015} derived from the RM to dispersion measure ratio
a mean line-of-sight magnetic field component of $ B_{\parallel} = 0.3
\pm 0.1 \mu$G and its variations across the 3C~196 field to $ 0.1
\mu$G. The 3C~196 field is well studied and we hope that follow-up
investigations can lead to more precise results.

The morphology of \hi filaments in the 3C~196 field is typical for the
CNM studied by \citet{Kalberla2016}. For ``unperturbed'' \hi, thus for
regions that are not affected by obvious CNM filaments, we find only
weak anisotropies around $Q_{\rm aver} \sim 2$. But this result, as well
as examples and results for strong anisotropies are based on a rather
limited sample.

This is to our knowledge the first study of turbulent anisotropies in
the \hi gas.  Some more investigations are necessary for a
generalization of our results, but the EBHIS
\citep{Winkel2016a,Winkel2016b}, 
as well as the GASS III \citep{Kalberla2015}, 
are available for such an analysis. The only limitation of these surveys
is caused by the different beam sizes, because of different telescope
diameters.  That disables an analysis of fields close to the ecliptic
equator. The LDS, as well as the Leiden/Argentine/Bonn survey
\citep[LAB,][]{Kalberla2005} are obsolete and should no longer be used to
determine properties of a turbulent flow. Our results are incompatible
with previous investigations using the LDS.

   \begin{acknowledgements}
     We thank F. Boulanger for providing data on the magnetic field
     direction in the FN1 region and V. Jeli{\'c} for comments. 
     The authors thank the Deutsche Forschungsgemeinschaft (DFG) for
     support under grant numbers KE757/7--1, KE757/7--2, KE757/7--3 and
     KE757/11--1. This research has made use of NASA’s Astrophysics Data
     System.  EBHIS is based on observations with the 100-m telescope of
     the MPIfR (Max-Planck-Institut f\"ur Radioastronomie) at Effelsberg.
   \end{acknowledgements}

\end{document}